\pdfoutput=1
\documentclass[
reprint, 
prd, 
aps, 
nofootinbib, 
showpacs, 
superscriptaddress]{revtex4-2}

\usepackage{graphicx}
\usepackage[usenames,dvipsnames]{xcolor}
\usepackage[utf8]{inputenc}
\usepackage[T1]{fontenc}
\usepackage{anyfontsize}
\usepackage{adjustbox}
\usepackage{placeins}
\usepackage{lipsum}
\usepackage{latexsym}
\usepackage{rotating}
\usepackage{mathtools}
\usepackage{amsmath, amssymb, amsthm, amsfonts}
\usepackage{mathrsfs}
\usepackage{bm}
\usepackage{enumerate}
\usepackage{soul}
\usepackage{natbib}
\usepackage{hyperref}
\usepackage{verbatim}
\usepackage[normalem]{ulem}
\usepackage{cancel}

\hypersetup{colorlinks=true,citecolor=NavyBlue,linkcolor=NavyBlue,urlcolor=NavyBlue}

\newcommand{\2}{\hspace{2mm}}
\newcommand{\N}{\textbf{N}}
\newcommand{\h}{\bar{h}}
\newcommand{\A}{\mathcal{A}}
\newcommand{\B}{\mathcal{B}}
\newcommand{\C}{\mathcal{C}}
\newcommand{\D}{\mathcal{D}}
\newcommand{{\KG}}{\mbox{\tiny{KG}}}
\newcommand{{\EA}}{\mbox{\tiny{EA}}}
\newcommand{\Scalar}{\mbox{\tiny{S}}}
\newcommand{\Vector}{\mbox{\tiny{V}}}
\newcommand{\Tensor}{\mbox{\tiny{T}}}
\newcommand{\TT}{\mbox{\tiny{TT}}}
\newcommand{\bb}{\mbox{\tiny{b}}}
\newcommand{\LL}{\mbox{\tiny{L}}}
\newcommand{\XX}{\mbox{\tiny{X}}}
\newcommand{\YY}{\mbox{\tiny{Y}}}
\newcommand{\edit}[1]{{\textcolor{black}{{{#1}}}}}

\numberwithin{equation}{section}
\renewcommand{\theequation}{\arabic{section}.\arabic{equation}}

\begin{document}

\title{Gravitational wave polarizations with different propagation speeds}

\author{Kristen Schumacher}
\affiliation{Department of Physics and Illinois Center for Advanced Studies of the Universe, University of Illinois at Urbana-Champaign, Urbana, IL, 61801, USA}

\author{Nicol\'as Yunes}
\affiliation{Department of Physics and Illinois Center for Advanced Studies of the Universe, University of Illinois at Urbana-Champaign, Urbana, IL, 61801, USA}

\author{Kent Yagi}
\affiliation{Department of Physics, University of Virginia, Charlottesville, Virginia 22904, USA}

\date{November 16, 2023}

\begin{abstract}
In some modified theories of gravity, gravitational waves can contain up to six different polarizations, which can travel at speeds different from that of light. 
Searches for these different polarizations in gravitational wave data are important because any detection would be clear evidence of new physics, while clear non-detections could constrain some modified theories. 
The first step toward searching the data for such gravitational wave content is the calculation of the amplitudes of these different polarizations. 
Here we present a model-independent method to obtain the different polarizations of gravitational waves directly from the metric perturbation in theories where these polarizations are allowed to travel at different speeds. 
We develop our calculations so that the same procedure works with either the metric perturbation itself or its trace-reversed form. 
Our results are in agreement with previous work in the limit that all polarization speeds are the speed of light. 
We demonstrate how our model-independent method can be used with two specific modified theories of gravity, suggesting its wide applicability to other theories that allow for different gravitational wave propagation speeds. 
We further extend the parameterized post-Einsteinian formalism to apply to such theories that travel with different speeds. Finally, we discuss how the different speeds of different polarizations may affect null stream tests of general relativity with gravitational wave observations by multiple interferometers. Differences in propagation speeds may make null streams ineffective or lead to the detection of what seem to be isolated scalar or vector modes.
\end{abstract}

\maketitle

\section{Introduction}
Over the course of the last decade, advancements in gravitational wave (GW) detection have made it possible to test gravity in unprecedented ways. 
These new studies are essential because, although Einstein's theory of general relativity (GR) has so far passed every possible test, hints exist that suggest it might need to be extended~\cite{Will:2014kxa, Clifton:2011jh}. 
Tests of GR with GW data~\cite{LIGOScientific:2016lio, Yunes:2016jcc, LIGOScientific:2018dkp, Berti:2018cxi, Berti:2018vdi, LIGOScientific:2019fpa, LIGOScientific:2020tif, LIGOScientific:2021sio} have made use of over 90 GW events from compact binary coalescences so far detected by the LIGO/Virgo collaboration~\cite{LIGOScientific:2021djp}. 
This number of detections is expected to increase dramatically in the future~\cite{KAGRA:2013rdx}, for instance, during the fourth-observing run that began recently. 
The abundance of new data promises to make it possible to learn even more about fundamental physics from the structure of GWs.

In GR, GWs contain only two polarizations, commonly known as the plus and cross modes. However, in some alternative theories of gravity, GWs may contain up to six modes and these modes may propagate at different speeds. Thus, any observation of additional modes would be indisputable evidence for new physics. To search for or try to constrain these additional polarizations with GW data, we need to be able to compute each polarization mode explicitly. From there, one can construct the response function in a GW detector and build a theory-specific waveform template to search for signals of these modified theories in GW data. 

One concrete example of a theory with additional GW polarizations that travel at different speeds is Einstein-\ae{}ther gravity. 
This theory is the most general Lorentz-violating model that can be constructed with a single additional unit vector field and that still leads to second-order field equations~\cite{Jacobson:2007veq}. 
Lorentz-violation is well motivated by attempts to quantize gravity~\cite{Mattingly:2005re}, and there are large regions of the possible parameter space in Einstein-\ae{}ther theory that have not yet been stringently constrained~\cite{Sarbach:2019yso, Oost:2018tcv, Schumacher:2023cxh}.
Attempts to constrain this theory with GW data are an active area of research~\cite{Schumacher:2023cxh}.

To constrain modified theories of gravity with GW data we need to compute each of the possible GW polarizations. 
In this work, we develop a simple method to do so, following the example of \cite{Chatziioannou:2012rf} and extending it to theories where the polarizations are allowed to travel at speeds different from that of light\footnote{While the tensor polarizations of GWs have been well constrained to propagate at the speed of light by the GW170817 event~\cite{LIGOScientific:2017zic}, other polarizations have neither been detected nor constrained to travel at a particular speed.}. 
We first compute the different GW polarizations from the linearized Riemann tensor~\cite{PWGravity2014}, keeping the speeds of the different polarizations free rather than fixing them to the speed of light.
Next, we develop an alternative approach that uses effective field theory ideas, following the example of~\cite{Chatziioannou:2012rf}, but extending the latter to the more general case of polarizations with different speeds.

We then validate our results by deriving the GW polarizations in specific modified theories of gravity. 
We begin with Einstein-\ae{}ther theory, comparing the results obtained here with those computed previously~\cite{Zhang:2019iim, Schumacher:2023cxh}. 
Next, we compute the different polarizations of GWs in khronometric gravity, a limiting case of Einstein-\ae{}ther theory that is consistent with the low-energy limit of Ho\v{r}ava gravity~\cite{Horava:2009uw}. Ho\v{r}ava gravity is a potential candidate for an ultraviolet completion of Einstein's GR~\cite{Horava:2009uw}. In both cases, we find that our method results in the correct GW polarizations in these modified theories, showing the robustness of the method. 

We next consider the well-established parameterized post-Einsteinian (ppE) framework of ~\cite{Yunes:2009ke}, which can be used to characterize deviations from GR in waveform templates. This framework was extended to theories with up to six polarizations by \cite{Chatziioannou:2012rf}. Here, we further extend it to theories where those polarizations can travel at different speeds. This generalization includes two new ppE parameters that correspond to the speeds of the scalar and vector polarizations. Finally, we examine how different speeds may affect the construction of null stream tests of general relativity with GWs detected by multiple interferometers. First considered by~\cite{Guersel:1989th} and~\cite{Chatterji:2006nh}, null streams are directions orthogonal to the plus and cross polarizations of GR. If GR is the correct description of nature, these streams would not contain any GW signal for a particular sky location. Reference \cite{Chatziioannou:2012rf} described a process by which one can use these null streams to constrain the existence of other polarizations with multiple detectors. We here discuss how the speed of the different polarizations affects the arrival time of a GW signal in these null streams and consider how that may affect future constraints. For instance, \textit{if the speeds of the polarizations are different by even a small amount ($\mathcal{O}(10^{-13})$) any beyond GR polarizations would be outside the standard detection window}, so null streams could not conclusively rule them out. Conversely, different propagation speeds could lead to the detection of  scalar or vector modes alone, because they may arrive much earlier than the corresponding tensor modes.

The main result of this work is a generalization of the method introduced in \cite{Chatziioannou:2012rf}, the latter of which has already been used to successfully compute GW polarizations in several modified theories of gravity (such as Brans-Dicke theory, Rosen's theory, and Lightman-Lee theory~\cite{Chatziioannou:2012rf}.).  
This generalization renders the method applicable to a wider class of theories, instead of only those in which all GW polarizations travel at the same speed. 
Tests of polarization content depend on having multiple detectors with multiple different orientations so that each detector observes different linear combinations of the polarization modes~\cite{Yunes:2013dva, Isi:2017fbj, Chatziioannou:2021mij}.
As more detectors come online and more events are observed, the potential for detection or constraints on any additional polarizations in the GW increases. 
Detection of additional polarizations would be clear evidence of physics beyond GR. 
Thus, our generalization will become increasingly important in the coming years, as more GW detectors are added to the global network and as we are able to probe the existence of additional polarizations and their propagation speeds. 

The remainder of this paper is organized as follows. 
In Sec.~\ref{sec:metricPerturbation}, we derive each polarization mode with different speeds directly from the metric perturbation. 
This is first done by computing the linearized Riemann tensor and then with an effective field theory approach. 
The result is compared to previous work. Then, in Sec.~\ref{sec:applicationToTheories}, we use our method to compute the different GW polarizations in two different modified theories of gravity, Einstein-\ae{}ther theory and khronometric gravity.
Section~\ref{sec:ppE} extends the ppE formalism to account for polarizations which travel at different speeds, while Sec.~\ref{sec:null_streams} discusses the impact of different speeds on null stream constraints. 
Section~\ref{sec:conclusions} summarizes our results and discusses potential applications. 
In the appendix ~\ref{sec:TRmetricPerturbation}, the process is repeated for the trace reversed metric perturbation. 
We compare this result to previous work, and to the result of Sec.~\ref{sec:metricPerturbation} in the case where all of the polarizations travel at the same speed, the speed of light. 
In the appendix~\ref{sec:appendixSpeeds}, we provide further detail about the difference in arrival times for modes that travel at different speeds.
Throughout this work, we use the following conventions: Greek letters in index lists specify spacetime coordinates, while Latin letters specify spatial coordinates only; parentheses in index lists stand for symmetrization, the Einstein summation convention is used, the metric signature is $(-,+,+,+)$, and $G = 1$ units are assumed; factors of $c$ are explicitly kept because of the different polarization speeds involved in this calculation. 

\section{Polarization Modes with Different Speeds From the Metric Perturbation} 
\label{sec:metricPerturbation}
In this section, we derive the GW polarizations when they are allowed to travel at speeds different from that of light. 
This derivation is first done by computing the linearized Riemann tensor with the metric perturbation. Then a more concise approach is developed, so that these polarizations can be calculated directly from the metric perturbation.

\subsection{Linearized Riemann Tensor Approach}
The GW polarizations can be read from their interaction with a detector. Such an interaction is determined by the geodesic deviation equation,  
\begin{equation}
\frac{d^2 \xi_j}{dt^2} = -c^2 R_{0j0k} \xi^k,
\end{equation}
where $\xi^\alpha$ is the spacetime vector separating the two test masses in the detector and $c$ is the speed of light \cite{PWGravity2014}. The Riemann tensor linearized about a flat Minkowski background is given by 
\begin{equation}
R_{0j0k} = -\frac{1}{2} \left(\partial_{00} h_{jk} + \partial_{jk} h_{00} -\partial_{0j} h_{0k} - \partial_{0k} h_{0j}\right),
\label{eqn:Riemann}
\end{equation}
where $h_{\alpha\beta}:= g_{\alpha\beta} - \eta_{\alpha\beta}$ is the metric perturbation, which can be broken up into the components  
\begin{subequations} \label{eqn:hdecomp}
\begin{align}
h_{00} &= \frac{\C}{c^4 R} , \label{eqn:h00} \\
h_{0j} &= \frac{\D_j}{c^4 R}, \label{eqn:h0j} \\
h_{jk} &= \frac{\A_{jk}}{c^4 R} \label{eqn:hjk}\,,
\end{align}
\end{subequations}
with $R$ the distance between the source and the detector and $\C$, $\D_{j}$ and $\A_{jk}$ scalar, (purely spatial) vector and (purely spatial, rank-2) tensor fields. In light of this decomposition, $h:= \eta^{\alpha\beta} h_{\alpha\beta} = -h_{00} + h_{kk} = (-\C + \A_{jk} \edit{\delta}^{jk})/(c^{4} R)$, with indices on the perturbations raised or lowered with the Minkowski metric. For ease of notation, the factor of $1/(c^4 R)$ will be absorbed into the functions $\C, \D^j$, and $\A^{jk}$ henceforth.

Following for example~\cite{PWGravity2014}, these expressions can be rewritten into their irreducible decompositions. More specifically, we use the tensorial nature of $\D^j$ and $\A_{jk}$ to uniquely decompose them into 
\begin{align}
\D_j &= \partial_j \D + \D_j^{\Tensor}\,,
\nonumber \\
\A_{jk} &= \frac{1}{3} \delta_{jk} \A + \left(\partial_{jk} - \frac{1}{3}\delta_{jk} \nabla^2 \right)\B + 2 \partial_{(j} \A_{k)}^{\Tensor} + \A_{jk}^{\TT}\,,
\end{align}
where $\D_j^{\Tensor}$ and $\A_j^{\Tensor}$ are transverse vector fields, i.e.~$\partial^{\5j} \D_j^{\Tensor} =0=\partial^{\5j} \A_j^{\Tensor}$, $\D$ and $\A$ are scalar fields, $\A_{jk}^{\TT}$ is a transverse and traceless tensor field, i.e.~$\partial^{\5j} \A_{jk}^{\TT} = 0$ and $\delta^{jk} \A_{jk}^{\TT} = 0$, and $\delta _{ij}$ is the spatial part of the Minkowski metric. 

Each of the functions in the fields that we have decomposed $h_{\alpha\beta}$ into depends on $\N$, a unit 3-vector aimed from the source to the detector, and on retarded time, $\tau:= t - R/v$, where $v$ is the velocity that the given field propagates at. This retarded time will now be given a subscript, S, V, or T, to denote whether it is associated with a scalar, a vector, or a tensor field respectively. This distinction is necessary since scalar, vector, and tensor polarizations in modified theories of gravity can, in principle, be allowed to travel with different speeds ($c_{\Scalar}, c_{\Vector}, c_{\Tensor}$). Retarded time then depends on this propagation speed through  
\begin{equation}
\tau_{\Scalar,\Vector,\Tensor} := t - R/c_{\Scalar,\Vector,\Tensor}. \label{eqn:retardedTime}
\end{equation}

For what follows, it will be useful to simplify the spatial derivatives of $h_{\alpha\beta}$. This is possible because $h_{\alpha\beta}$ depends on spatial coordinates through $\tau,$ $\N$ and $R^{-1}$. Thus, taking spatial derivatives of these, one finds
\begin{subequations}
\begin{align}
\partial_j \tau_{\Scalar,\Vector,\Tensor} &= -c_{\Scalar,\Vector,\Tensor}^{-1} N_j, \\
\partial_j N^k &= \mathcal{O}(R^{-1}), \\
\partial_j R^{-1} &= \mathcal{O}(R^{-2}). 
\end{align}
\end{subequations}
Ignoring terms of higher order in $1/R$, the only dependence that matters for the spatial derivative is that on $\tau_{\Scalar,\Vector,\Tensor}$. Thus, every spatial derivative of $h^{\alpha\beta}$ can be rewritten as a time derivative because $\partial_t = \partial_{\tau}$. For example, $\partial_j \D(\tau_{\Scalar}, \N) = -c_{\Scalar}^{-1} N_j \partial_t \D + {\cal{O}}(R^{-1})$. We will also introduce the notation $x^{0} = c t$, so that $\partial_0 = c^{-1} \partial_t$ to simplify notation.

Using the derivative relation, the vector and tensor decompositions (with some renaming so that the negative signs and $\partial_t$ are not shown for simplicity) of $h_{\alpha\beta}$ can be rewritten as 
\begin{subequations} \label{eqn:gravitationalPotentials}
\begin{align}
h_{00} &= \C(\tau_{\Scalar}, \N), \\
h_{0j} &= \frac{1}{c_{\Scalar}} N_j \D(\tau_{\Scalar}, \N) + \D_j^{\Tensor} (\tau_{\Vector}, \N), \\
h_{jk} &= \frac{\delta_{jk}}{3} \A(\tau_{\Scalar}, \N) + \frac{1}{c_{\Scalar}^2} \left(N_j N_k - \frac{\delta_{jk}}{3} \right)\B(\tau_{\Scalar}, \N) \nonumber \\
&\2\2+\frac{1}{c_{\Vector}} N_j \A_k^{\Tensor}(\tau_{\Vector},\N) + \frac{1}{c_{\Vector}} N_k \A_j^{\Tensor}(\tau_{\Vector}, N) \nonumber \\
&\2\2 + \A_{jk}^{\TT}(\tau_{\Tensor}, \N),
\end{align}
\end{subequations}
where we have explicitly included the dependence of these fields on $\tau$ and $\N$.

Inserting this decomposition into the linearized Riemann tensor (Eq.~\eqref{eqn:Riemann}) and simplifying with the derivative rules developed above yields 
\begin{align}
R_{0j0k} &= -\frac{1}{2c^2} \partial_{tt} \left[ \frac{\delta_{jk}}{3} \A + \frac{1}{c_{\Scalar}^2} \left(N_j N_k - \frac{\delta_{jk}}{3} \right) \B\right. \nonumber \\
&\2\2+ \frac{1}{c_{\Vector}} N_j \A^T_k + \frac{1}{c_{\Vector}} N_k \A^T_j + \A_{jk}^{\TT} \nonumber \\
&\2\2 + \frac{c^2}{c_{\Scalar}^2} N_j N_k \C + \frac{2c}{c_{\Scalar}^2} N_j N_k \D  \nonumber\\
&\2\2\left.+ \frac{c}{c_{\Vector}} N_j \D_k^T + \frac{c}{c_{\Vector}} N_k \D_j^T \right]\,.
\end{align}
This expression can be rearranged into the form 
\begin{multline}
R_{0j0k} = -\frac{1}{2c^2} \partial_{tt} \Big[(\delta_{jk} - N_j N_k) A^{\bb} + N_j N_k A^{\LL}  \\
\left. + 2N_{(j} A_{k)}^{\Vector} + A_{jk}^{\TT}\right]\,,
\end{multline}
where 
\begin{subequations}\label{eqn:results}
\begin{align}
A^{\bb} &:= \frac{1}{3}\left[\A - \frac{1}{c_{\Scalar}^2} \B \right], \label{eqn:breathingMode}\\
A^{\LL} &:= \frac{1}{3} \A + \frac{2}{3c_{\Scalar}^2} \B  + \frac{c^2}{c_{\Scalar}^2}\C + 2\frac{c}{c_{\Scalar}^2} \D, \label{eqn:longitudinalMode}\\
A^{\Vector}_k &:= \frac{1}{c_{\Vector}} \left[\A_k^{\Tensor} + c\D_k^{\Tensor}\right]. \label{eqn:vectorialMode}
\end{align}
\end{subequations}
These are the scalar breathing mode, $A^{\bb}$, and the scalar longitudinal mode, $A^{\LL}$. The quantity $A^{\Vector}_k$ contains 2 vector modes (because the transverse condition eliminates 1 degree of freedom). These polarization modes, together with $A_{jk}^{\TT}$, which contains two tensor modes (because the transverse-traceless condition eliminates 3 degrees of freedom), make up the 6 possible GW polarization modes in modified gravity theories. The results presented above can be recast in terms of the irreducible decomposition of the trace-reversed metric perturbation (instead of using the standard metric perturbation itself), as we do in Appendix~\ref{sec:TRmetricPerturbation}, Eq.~\eqref{eqn:TRresults}.

\subsection{An Effective Field Theory Approach} \label{sec:RegularPerturbationResult}
Alternatively, we can construct operators that act on the metric perturbation directly to obtain the polarization modes. For instance, to derive the scalar modes, consider the four ways to produce a scalar from the gravitational potentials: $h_{00},$ $N^j h_{0j},$ $N^j N^k h_{jk}$ and $\delta^{jk} h_{jk}$. The scalar breathing mode $A^{\bb}$ will be some linear combination of these, 
\begin{equation}
A^{\bb} = a_1 h_{00} + a_2 N^j h_{0j} + a_3 N^j N^k h_{jk} + a_4 \delta^{jk} h_{jk}\,,
\end{equation}
with constants $a_1, a_2, a_3$, and $a_4$, such that the linear combination matches the $A^{\bb}$ found previously in Eq.~\eqref{eqn:breathingMode}. Inserting Eq.~\eqref{eqn:gravitationalPotentials} into this linear combination and simplifying leads to 
\begin{equation}
A^{\bb} = a_1 \C + \frac{a_2}{c_{\Scalar}} \D + a_3 \left(\frac{1}{3} \A + \frac{2}{3c_{\Scalar}^2} \B\right) + a_4 \A.
\end{equation}
Comparing with Eq.~\eqref{eqn:breathingMode} and matching coefficients gives 
\begin{equation}
a_1 = 0,\2 a_2 = 0,\2 a_3 = -\frac{1}{2},\2 a_4 = \frac{1}{2}. 
\end{equation}
Therefore, 
\begin{equation}
A^{\bb} = \frac{1}{2} \left(\delta^{jk} h_{jk} - N^j N^k h_{jk}\right).
\end{equation}
Following the same procedure for the scalar longitudinal mode, we arrive at 
\begin{equation}
A^{\LL} = \frac{c^2}{c_{\Scalar}^2} h_{00} + 2\frac{c}{c_{\Scalar}} N^j h_{0j}+ N^j N^k h_{jk}.
\end{equation}

In order to derive similar expressions for the vector modes, we consider the two ways to construct a vector from the gravitational potentials: $h_{0j}, N^i h_{ij}$. Since this vector should be transverse (because the vector polarizations are transverse), the transverse projector, $P^j_{\1k} = \delta^j_k - N^j N_k$ \cite{PWGravity2014}, 
should also be used to remove longitudinal components. Thus, we study the linear combination 
\begin{equation}
A^{\Vector}_k = P^j_{\1k} \left(a_5 h_{0j} + a_6 N^i h_{ij}\right),
\end{equation}
with some constants $a_5$ and $a_6$, such that this linear combination matches the $A^{\Vector}_k$ found previously in Eq.~\eqref{eqn:vectorialMode}. Inserting Eq.~\eqref{eqn:gravitationalPotentials} into this linear combination and simplifying yields
\begin{equation}
A^{\Vector}_k = a_5 \D\edit{_k^{\Tensor}} + a_6 \frac{1}{c_{\Vector}} \A\edit{_k^{\Tensor}}.
\end{equation}
Matching coefficients with Eq.~\eqref{eqn:vectorialMode} gives
\begin{align}
a_5 = \frac{c}{c_{\Vector}},\2 a_6 = 1.
\end{align}
Thus, 
\begin{equation}
A^{\Vector}_k = P^j_{\1k} \left(N^i h_{ij} + \frac{c}{c_{\Vector}} h_{0j}\right). 
\end{equation}

In summary, the polarization modes can be extracted directly from the metric perturbation through
\begin{subequations} \label{eqn:resultsDiffSpeeds}
\begin{align}
A^{\bb} &= \frac{1}{2} \left(\delta^{jk} h_{jk} - N^j N^k h_{jk}\right), \label{eqn:breathingModeDiffSpeeds}\\
A^{\LL} &= \frac{c^2}{c_{\Scalar}^2} h_{00} + 2\frac{c}{c_{\Scalar}} N^j h_{0j}+ N^j N^k h_{jk}, \label{eqn:longitudinalModeDiffSpeeds}\\
A^{\Vector}_k &= P^j_{\1k} \left(N^i h_{ij} + \frac{c}{c_{\Vector}} h_{0j}\right)\,, \label{eqn:vectorialModeDiffSpeeds} \\
A^{\TT}_{jk} &= {\rm{TT}}^{lm}{}_{jk} h_{lm}\,,
\end{align}
\end{subequations}
where we have included the transverse-traceless mode through the transverse-traceless operator ${\rm{TT}}^{lm}{}_{jk} = P^{l}_{\1j} P^{m}_{\2k} - \frac{1}{2} P^{lm} P_{jk}$~\cite{PWGravity2014}.
These expressions, calculated directly from the metric perturbation, are consistent with what was found from the linearized Riemann tensor calculation (Eq.~\eqref{eqn:results}). That is, if we insert Eq.~\eqref{eqn:gravitationalPotentials} into the above expressions, we recover Eq.~\eqref{eqn:results} exactly. Observe that these expressions depend  explicitly on $c, c_{\Scalar}$, and $c_{\Vector}$, but are independent of $c_{\Tensor}$. The above results can again be recast in terms of the irreducible decomposition of the trace-reversed metric perturbation, as we do in Appendix~\ref{sec:TRmetricPerturbation}.

\subsection{Comparison to previous results} 
Let us take the limit where all modes travel at the same speed, namely the speed of light. If $c_{\Scalar,\Vector,\Tensor} \rightarrow c$, then the expressions of Eq.~\eqref{eqn:resultsDiffSpeeds} reduce to
\begin{subequations} \label{eqn:limitSameSpeeds}
\begin{align}
A\edit{^{\bb}} &= \frac{1}{2} \left(\delta^{jk}h_{jk} - N^j N^k h_{jk}\right), \label{eqn:breathingModeSameSpeeds}\\
A\edit{^{\LL}} &= h_{00} + 2N^j h_{0j} + N^j N^k h_{jk}, \label{eqn:longitudinalModeSameSpeeds} \\
A^{\Vector}_k &= P^j_{\1k} \left(N^i h_{ij} + h_{0j} \right). \label{eqn:vectorialModeSameSpeeds}
\end{align}
\end{subequations}
We convert these expressions to ones that act on the trace-reversed metric perturbation so that they can be directly compared with results from Eqs.~(9a)--(9c) in Chatziioannou \emph{et. al}~\cite{Chatziioannou:2012rf}. Using the definition for the trace-reversed metric perturbation $\h_{\alpha\beta}$,  
\begin{equation}
h_{\alpha\beta} = \h_{\alpha\beta} - \frac{1}{2}\eta_{\alpha\beta} \h  \label{eqn:metricConversionRule}
\end{equation}
where $\h = \eta^{\alpha\beta} \h_{\alpha\beta} = -\h_{00} + \h_{kk} = -h = \eta^{\alpha \beta}h_{\alpha \beta}$, the expressions in Eq.~\eqref{eqn:limitSameSpeeds} become 
\begin{align*}
\label{eqn:finalResultSameSpeeds}
A^{\bb} &= \frac{1}{2} \left(\h_{00} - N^j N^k \h_{jk}\right), \\
A^{\LL} &= \h_{00} + 2N^j \h_{0j} + N^j N^k \h_{jk}, \\
A^{\Vector}_k &= P^j_{\1k} \left(N^i \h_{ij} + \h_{0j} \right). 
\end{align*}
This expression matches identically that in Eqs. (9a)--(9c) of Chatziiouannou \emph{et. al}~\cite{Chatziioannou:2012rf}, after recognizing that $\bar{h}_{00} = \bar{h}^{00}$, $\bar{h}_{jk} = \bar{h}^{jk}$, $\bar{h}_{0i} = -\bar{h}^{0i}$, and finding an overall minus sign typo in the $A_{\bb}$ expression of ~\cite{Chatziioannou:2012rf}.

\section{Application to Different Theories}
\label{sec:applicationToTheories}
In this section, we use our generic results to compute the different GW polarizations in two specific modified theories of gravity. First, we perform this calculation in Einstein-\ae{}ther theory and we compare the result to previous work~\cite{Zhang:2019iim}. Then, we repeat the procedure with khronometric gravity, comparing these polarizations to the limiting case of Einstein-\ae{}ther theory and updating the work of~\cite{Hansen:2014ewa}. Note that for this section \textit{and this section only} we set $c = 1$ to be consistent with the notation of references that provide the metric perturbation and equations of motion for these theories.

\subsection{Einstein-\ae{}ther Theory} 
In order to validate the result of Eq. \eqref{eqn:resultsDiffSpeeds} further, we can use it to compute the GW polarizations in a particular theory of gravity and then compare the result to GW polarizations calculated independently in other existing works. For instance, according to \cite{Zhang:2019iim}, the metric perturbation of Einstein-\ae{}ther theory can be decomposed into  
\begin{subequations} \label{eqn:AEmetricpert}
\begin{align}
h_{00}^{\EA} &= 2w^0, \\
h_{0i}^{\EA} &= \gamma_i + \partial_i \gamma, \\
h_{ij}^{\EA} &= \partial_{ij} \phi^{\EA} + \frac{1}{2} \left(\delta_{ij} \Delta - \partial_{ij} \right) f^{\EA}  \nonumber \\
&\2\2+ \partial_j \phi_i^{\EA} + \partial_i \phi_j^{\EA} + \phi_{ij}^{\EA}  ,
\end{align}
\end{subequations}
where $\Delta \equiv \delta^{ij} \partial_i \partial_j$\edit{, $w^0 = u^0 - 1$ and $w^i = u^i$ where $u^\alpha$ is the four-velocity of the \ae{}ther field, } and 
\begin{align*}
\partial^i \gamma_i = \partial^i \phi_i^{\EA} &= 0, \\
 \partial^j \phi_{ij}^{\EA} = 0, \2 \delta^{ij} \phi_{ij}^{\EA} &= 0.
\end{align*}
Inserting this metric perturbation (Eq.~\eqref{eqn:AEmetricpert}) into the results of Sec.~\ref{sec:RegularPerturbationResult} (Eq.~\eqref{eqn:resultsDiffSpeeds}) and using the previously-developed derivative rule ($\partial_j = -{c_{\Scalar,\Vector,\Tensor}}^{-1} N_j \partial_t$) yields 
\begin{subequations} \label{eqn:AEresults}
\begin{align}
A^{\bb} &= \frac{1}{2} \Delta f^{\EA}, \\
A^{\LL} &= \frac{1}{2} \Delta f^{\EA} - \frac{1}{2c_{\Scalar}^2} \ddot{f}^{\EA} - \frac{1}{c_{\Scalar}^2} 2 \dot{\gamma} + \frac{1}{c_{\Scalar}^2} (2w^0)  + \frac{1}{c_{\Scalar}^2} \ddot{\phi}^{\EA}, \\
A^{\Vector}_i &= - \frac{1}{c_{\Vector}} \dot{\phi}_i^{\EA} + \frac{1}{c_{\Vector}} \gamma_i.
\end{align}
\end{subequations}
These are the GW polarizations that our general procedure predicts in Einstein-\ae{}ther theory. The fields $f^{{\EA}}$, $\gamma$, $w^{0}$, $\gamma_{i}$, $\phi_{i}^{{\EA}}$, and $\phi^{{\EA}}$ satisfy the linearized field equations in that theory, which must be solved for a given physical system to obtain explicit functions of time that can then be compared to GW data. The dot notation here represents time derivatives, and $(c_{\Scalar}^2)^{-1} \ddot{f}_{{\EA}}$ is equivalent to $\Delta f_{{\EA}}$ by the definition of $\Delta$ and the derivative rule. 

More work needs to be done to show that these polarizations match the results of \cite{Zhang:2019iim}. From~\cite{Lin:2018ken}, we know that the Einstein-\ae{}ther linearized field equations require that
\begin{align*}
c_{13} (\nu_i + \dot{\phi}_i^{\EA}) + \gamma_i - \dot{\phi}_i ^{\EA}&= 0, \\
F^{\EA} - c_{14}(2w^0) + 2c_{14} (\dot{\gamma} + \dot{\nu}) &+ 0, \\
(1 + c_2) \dot{f}^{\EA} + c_{123}(\dot{\phi}^{\EA} + 2\nu) &= 0,
\end{align*}
where \edit{$w_i = \nu_i + \partial_i \nu$ }, $F^{\EA} \equiv \Delta f^{\EA}$, $\{c_1, c_2, c_3, c_4\}$ are coupling constants, and $c_{ij} = c_i + c_j$ while $c_{ijk} = c_i + c_j + c_k$ \cite{Lin:2018ken}. These expressions directly imply 
\begin{subequations} \label{eqn:conditions}
\begin{align}
\gamma_i &= (1 - c_{13}) \dot{\phi}_i^{\EA} - c_{13} \nu_i\,, \\
2w^0 &= \frac{1}{c_{14}} F^{\EA} + 2\dot{\gamma} + 2\dot{\nu}\,, \\
\dot{\phi}^{\EA} &= -\frac{(1+c_2)}{c_{123}} \dot{f}^{\EA} - 2\nu\,.
\end{align}
\end{subequations}
Furthermore, the Einstein-\ae{}ther field equations also require that the scalar modes in this theory travel at a speed~\cite{Jacobson:2007veq,Mattingly:2005re,Zhang:2019iim} 
\begin{equation} \label{eqn:cSAE}
c_{\Scalar}^2 = \frac{(2-c_{14})c_{123}}{(2+3c_2+c_{13})(1-c_{13})c_{14}}\,.
\end{equation}
Using these relations (Eq.~\eqref{eqn:conditions} and Eq.~\eqref{eqn:cSAE}) we can simplify the expression for $A^{\LL}$ given in Eq.~\eqref{eqn:AEresults}: 
\begin{align}
A^{\LL} &=  \frac{1}{2} \Delta f^{\EA} - \frac{1}{2c_{\Scalar}^2} \ddot{f}^{\EA} - \frac{1}{c_{\Scalar}^2} 2 \dot{\gamma} + \frac{1}{c_{\Scalar}^2} (2w^0) + \frac{1}{c_{\Scalar}^2} \ddot{\phi}^{\EA} \nonumber\\
&= \frac{1}{2} F^{\EA} + \frac{1}{c_{\Scalar}^2 c_{14}} F^{\EA} - \frac{1}{2} F^{\EA} - \frac{(1+c_2)}{c_{123}} F^{\EA} \nonumber\\
&= \left( \frac{1}{2} + \frac{c_{14} - 2c_{13}}{2c_{\Scalar}^2 c_{14}(1-c_{13})} \right) F^{\EA} \nonumber\\
 &= (1 + 2\beta_2) A^{\bb},
\end{align}
where, as pointed out in~\cite{Schumacher:2023cxh}, 
\begin{equation}
\label{eq:beta2}
\beta_2 \equiv \frac{c_{14} - 2c_{13}}{2c_{14}(1 - c_{13})c_{\Scalar}^2}\,.
\end{equation} 
Simplifying the expression for $A^{\Vector}_i$ from Eq.~\eqref{eqn:AEresults} with the conditions in Eq.~\eqref{eqn:conditions} yields 
\begin{align*}
A^{\Vector}_i &= -\frac{c_{13}}{c_{\Vector}} \left(\nu_i + \dot{\phi}_i^{\EA} \right) = \frac{1}{2} \beta_1 \left(\nu_i + \dot{\phi}_i^{\EA} \right),
\end{align*}
where 
\begin{align}
\label{eq:beta1def}
\beta_1 \equiv -\frac{2c_{13}}{c_{\Vector}} .
\end{align}

Thus, to summarize, we have arrived at 
\begin{subequations} \label{eqn:resultsEA}
\begin{align}
A^{\bb} &= \frac{1}{2} F^{\EA}, \label{eqn:breathingModeEA}\\
A^{\LL} &= (1 + 2\beta_2) A^{\bb}, \label{eqn:longitudinalModeEA}\\
A^{\Vector}_i &= \frac{1}{2} \beta_1 \left(\nu_i + \dot{\phi}_i^{\EA}\right). \label{eqn:vectorModeEA}
\end{align}
\end{subequations}
Using the following relations from~\cite{Chatziioannou:2012rf}, 
\begin{align*}
h_{\bb} &= A^{\bb}, \2\2 h_{\LL} = A^{\LL}, \\
h_{\XX} &= e^i_{\XX} A^{\Vector}_i, \2\2 h_{\YY} = e^i_{\YY} A^{\Vector}_i, 
\end{align*} 
where $h_{\XX,\YY}$ are the two vector polarization modes, 
we can convert the scalar breathing mode, scalar longitudinal mode, and vector mode into 
\begin{subequations} \label{eqn:AEfinal}
\begin{align}
h_{\bb} &= \frac{1}{2} F^{\EA}, \\
h_{\LL} &= (1 + 2\beta_2) h_{\bb}, \\
h_{\XX} &= \frac{1}{2} \beta_1 \left(\nu_i + \dot{\phi}_i^{\EA}\right) e^i_{\XX}, \\
h_{\YY} &= \frac{1}{2} \beta_1 \left(\nu_i + \dot{\phi}_i^{\EA}\right) e^i_{\YY}.
\end{align}
\end{subequations}
Following the convention of \cite{Zhang:2019iim}, we choose a gauge, which sets $\phi_i = 0$ and $\nu = \gamma = 0$. 
Now upon comparison with Eq. (3.28) of \cite{Zhang:2019iim}, which listed 
\begin{align*}
h_{\bb} &= \frac{1}{2} F^{\EA},\2\2 h_{\LL} = (1 + 2\beta_2)h_{\bb}, \\
h_{\XX} &= \frac{1}{2} \beta_1 \nu^i e^i_{\XX},\2\2 h_{\YY}= \frac{1}{2} \beta_1 \nu^i e^i_{\YY},
\end{align*}
we find that our expressions match exactly.
Therefore, our generic procedure to extract the GW polarizations from the metric perturbation has been used to correctly recover the specific GW polarizations of Einstein-\ae{}ther theory computed independently elsewhere.

\subsection{Khronometric Gravity} 
In harmonic coordinates and in an appropriate gauge\footnote{Following the convention of Foster, a gauge is chosen such that $\nu = \gamma = \phi_i = 0$ \cite{Foster:2007gr}.}, the metric perturbation for khronometric gravity is given by \cite{Hansen:2014ewa}\footnote{That work was missing the $\Delta$ in $h_{00}$ and did not define the projection operator explicitly, so both have been fixed here.} 
\begin{subequations} \label{eqn:KGmetricpert}
\begin{align}
h_{00}^{{\KG}} &= \frac{1}{\alpha_{{\KG}} } \Delta f^{{\KG}}, \\
h_{0j}^{{\KG}} &= 0, \\
h_{jk}^{{\KG}} &= \partial_{jk} \phi^{\KG} + \frac{1}{2} \left(\delta_{jk} \Delta - \partial_j \partial_k \right) f^{\KG} + \phi_{jk}^{\KG}, 
\end{align}
\end{subequations}
where $\Delta \equiv \delta^{jk} \partial_j \partial_k$, $\phi_{jk}^{\KG}$ obeys the conditions 
\begin{equation} 
\partial^k \phi_{jk}^{\KG} = 0,\2\2 \delta^{jk} \phi_{jk}^{{\KG}} = 0\,, 
\end{equation}
and $\alpha_{{\KG}}$\edit{, $\beta_{{\KG}}$, and $\lambda_{{\KG}}$ are} parameter\edit{s} in the theory\edit{, which appear in the action}.
Inserting this metric perturbation (Eq.~\eqref{eqn:KGmetricpert}) into the results from Sec.~\ref{sec:RegularPerturbationResult} (Eq.~\eqref{eqn:resultsDiffSpeeds}) and using the previously-developed derivative rule $(\partial_j = - c_{\Scalar,\Vector,\Tensor}^{-1} N_j \partial_t)$, we find 
\begin{subequations}
\begin{align}
A^{\bb} &= \frac{1}{2} \Delta f^{\KG}\,, \label{eqn:breathingModeKG}\\
A^{\LL} &=  \Delta \phi^{\KG} +  \frac{1}{c_{\Scalar}^2\alpha_{\KG}} \Delta f^{\KG}\,,  \label{eqn:longitudinalModeKG}\\
A^{\Vector}_k &= 0\,. \label{eqn:vectorModeKG}
\end{align}
\end{subequations}
These are the GW polarizations that our general procedure predicts in khronometric gravity. The fields $f^{{\KG}}$ and $\phi^{\KG}$ satisfy the linearized field equations in that theory, which must be solved for a given physical system to obtain explicit functions of time that can then be compared to GW data.

As a check on this result, we can use the fact that khronometric gravity is a limiting case of Einstein-\ae{}ther theory to derive the GW polarizations in a different way using previous independent work. This limit is when $c_{-} = c_1 - c_3 \rightarrow \infty$, while the other parameters of Einstein-\ae{}ther theory are kept fixed. According to~\cite{Jacobson:2013xta}, this means that $c_+\edit{=c_1+c_3}, c_{14}$ and $c_2$ remain fixed while $c_1, c_3$ and $c_4$ diverge. In this limiting case, we can map the parameters of Einstein-\ae{}ther theory to those of khronometric gravity. Specifically, 
\begin{subequations}
\begin{align}
\alpha_{{\KG}} &= c_{14} = c_1 + c_4, \\
\beta_{{\KG}} &= c_+ = c_1 + c_3, \\
\lambda_{{\KG}} &= c_2.
\end{align}
\end{subequations}
In this limit and in the gauge we are working in ($\nu = \gamma = \phi_i = 0$), the equations of motion for Einstein-\ae{}ther theory (Eq.~\eqref{eqn:conditions}) become the khronometric equations
\begin{subequations}
\begin{align}
c_{13} \nu_i &= 0, \label{eqn:condition1} \\
h_{00}^{{\KG}} &= \frac{1}{\alpha_{\KG}} \Delta f^{\KG}, \\
\dot{\phi}^{\KG} &= - \frac{1 + \lambda_{\KG}}{\lambda_{\KG} + \beta_{\KG}} \dot{f}^{\KG}. \label{eqn:condition3}
\end{align}
\end{subequations}

Let us now recall that the polarizations in the Einstein-\ae{}ther case are given in Eq.~\eqref{eqn:resultsEA} and apply the limit above to derive the GW polarizations in khronometric gravity, starting with $A^{\bb}$ first.
Since this GW polarization (Eq.~\eqref{eqn:breathingModeEA}) does not contain any divergent quantity, taking the limit of this expression leaves it unchanged. Therefore, the breathing GW polarization in khronometric gravity is simply that of Eq.~\eqref{eqn:breathingModeEA} with the replacement $f^{\EA} \to f^{\KG}$. We observe then immediately that this is the same as what our general method predicted in Eq.~\eqref{eqn:breathingModeKG}.

Let us now consider $A^{\LL}$ (Eq.~\eqref{eqn:longitudinalModeEA}). Recalling the Einstein-\ae{}ther definition of $\beta_{2}$ from Eq.~\eqref{eq:beta2}, with $c_{\Scalar}$ given in Eq.~\eqref{eqn:cSAE}, we realize that only the fixed quantities ($c_2, c_{13}$ and $c_{14}$) appear in these expressions, and thus, it remains unchanged in the khronometric limit. We can therefore rewrite these expressions in terms of the khronometric parameters as
\begin{subequations}
\begin{align}
\beta_2 &= \frac{\alpha_{\KG} - 2\beta_{\KG}}{2\alpha_{\KG} (1 - \beta_{\KG}) c_{\Scalar}^2}, \\
c_{\Scalar}^2 &=  \frac{(2 - \alpha_{\KG})(\lambda_{\KG} + \beta_{\KG})}{(2+3\lambda_{\KG} + \beta_{\KG})(1-\beta_{\KG}) \alpha_{\KG}}.
\end{align}
\end{subequations}
With this in hand, the Einstein-\ae{}ther polarizations of Eq.~\eqref{eqn:longitudinalModeEA} become
\begin{align}
A^{\LL} &= \frac{1}{2} \Delta f^{\KG} + \frac{\alpha_{\KG} - 2\beta_{\KG}}{2\alpha_{\KG} (1 - \beta_{\KG}) c_{\Scalar}^2} \Delta f^{\KG} \nonumber \\
&= \frac{1}{2} \Delta f^{\KG} + \left(\frac{1}{c_{\Scalar}^2\alpha_{\KG}} \Delta f^{\KG} - \frac{1}{2} \Delta f^{\KG} \right. \nonumber \\
&\2\2 \left. - \frac{(1+\lambda_{\KG})}{(\lambda_{\KG} + \beta_{\KG})} \Delta f^{\KG} \right) \nonumber \\
&= \frac{1}{c_{\Scalar}^2 \alpha_{\KG}} \Delta f^{\KG} + \Delta \phi^{\KG}\,,
\end{align}
where Eq.~\eqref{eqn:condition3} was used to get to the last line. Thus, the longitudinal mode of Einstein-\ae{}ther theory (Eq.~\eqref{eqn:longitudinalModeEA}) in the khronometric limit yields the same polarizations as our general procedure applied to khronometric gravity (Eq.~\eqref{eqn:longitudinalModeKG}). 

Finally, the Einstein-\ae{}ther theory $A^{\Vector}_k$ (Eq.~\eqref{eqn:vectorModeEA}) vanishes in the khronometric limit because $\phi_k = 0 $ in this gauge and $\nu_k = 0$ by Eq.~\eqref{eqn:condition1}. Alternatively, we could arrive at the same result by realizing that~\cite{Zhang:2019iim}
\begin{equation}
c_{\Vector}^2 = \frac{2c_1 - c_{13} c_-}{2(1-c_{13})c_{14}} 
\end{equation}
diverges in the khronometric limit (because $c_- \rightarrow \infty$ and $c_1 \rightarrow \infty$), which then implies that $\beta_{1}$, as defined in Eq.~\eqref{eq:beta1def} must vanish. Once again, the Einstein-\ae{}ther result (Eq.~\eqref{eqn:vectorModeEA}) in the appropriate limit agrees with the vectorial mode of khronometric gravity derived with our generic method (Eq.~\eqref{eqn:vectorModeKG}). To reiterate, we find that the scalar breathing mode, the scalar longitudinal mode, and the vectorial mode derived in khronometric gravity with our general procedure agree with those of Einstein-\ae{}ther in the khronometric limit.

\section{Generalized ppE Formalism}
\label{sec:ppE}
The ppE framework can be used to characterize modifications or extensions to GR without specifying a particular theory. As such, it has been used extensively since its introduction~\cite{Yunes:2009ke, Yunes:2013dva, Will:2014kxa, Berti:2015itd, LISA:2022kgy}. The parameters of this framework describe how the amplitude and phase of the response function may change due to generic deviations from GR. The original formalism considered modifications to the two tensor polarizations of GR in a single detector for the $\ell=2$ orbital harmonic~\cite{Yunes:2009ke}. This was extended by~\cite{Chatziioannou:2012rf} to theories that have up to six polarizations. Here we briefly review the extensions made in~\cite{Chatziioannou:2012rf} and then introduce further modifications to account for modes that travel at different speeds.

As shown by~\cite{Yunes:2009ke}, the Fourier transform of the response function can be written as 
\begin{equation}
    \tilde{h}_{\text{ppE}} (f) = \mathcal{A}_{\text{GR}} u_2^{-7/2} (1 + \alpha u_2^a) e^{-i\Psi_{\text{GR}}^{(2)}} e^{i\beta u_2^b}\,. \label{eqn:ppE}
\end{equation}
In this expression, the variables $\{\alpha, a, \beta, b\}$ are the ppE parameters that generically modify the waveform, $u_2$ is a convenient combination of the mass and frequency, and $\Psi_{\text{GR}}^{(2)}$ and $\mathcal{A}_{\text{GR}}$ are the phase and frequency-independent amplitude in GR respectively. Explicitly,
\begin{align}  
    u_\ell &= \left( \frac{2\pi \mathcal{M}f}{\ell} \right)^{1/3}
    \,,\\
    \Psi_{\text{GR}}^{(\ell)} &= -2\pi f t_c + \ell \Phi_c + \frac{\pi}{4} \nonumber\\
    &\2\2- \frac{3\ell}{256 u_\ell^5} \sum_{n=0}^7 u_\ell^{n/3} (c_n^{\text{PN}} + l_n^{\text{PN}} \ln u)\,, 
\label{eq:Psi_GR}
\\
    \mathcal{A}_{\text{GR}} &= \left( \frac{5\pi}{96} \right)^{1/2} \frac{\mathcal{M}^2}{R} \left[ F_+^2 (1 + \cos^2\iota)^2 + 4 F_\times^2 \cos^2\iota \right]^{1/2}\,, 
\end{align}
where only the $\ell =2$ harmonic appears in Eq.~\eqref{eqn:ppE}, $c_n^{\text{PN}}$ and $l_n^{\text{PN}}$ are known post-Newtonian (PN) coefficients, given in~\cite{Buonanno:2009zt}\footnote{ Recently, the phase was computed up to 4.5PN order in GR and contains terms that are proportional to $(\ln u)^2$~\cite{Blanchet:2023bwj}. One could easily generalize Eq.~\eqref{eq:Psi_GR} to include these terms in $\Psi_{\text{GR}}$.}, and $F_+$ and $F_\times$ are angle pattern functions as defined in~\cite{PWGravity2014}.

Equation~\eqref{eqn:ppE} was first extended to include the $\ell =1$ harmonic since this harmonic is excited by theories which contain additional polarizations in the GWs. According to~\cite{Chatziioannou:2012rf}, assuming all polarizations travel at the speed of light, the new version for a single detector is 
\begin{align}
    \tilde{h}_{\text{ppE}}^{SD} (f) &= \mathcal{A}_{\text{GR}} u_2^{-7/2} (1 + \alpha u_2^a) e^{-i\Psi_{\text{GR}}^{(2)}} e^{i\beta u_2^b} \nonumber \\
    &\2\2 + \gamma u_1^c e^{-i\Psi_{\text{GR}}^{(1)}} e^{i\delta u_1^d}\,,
    \label{eqn:ppE_all_harmonics}
\end{align}
where $\{\gamma, c, \delta, d\}$ are the new ppE parameters necessary to describe modifications to the $\ell =1$ harmonic. To further generalize this to all six polarizations the number of parameters was increased so that there is one for each possible polarization mode for each harmonic~\cite{Chatziioannou:2012rf}\footnote{This is Eq.~(145) of that work and not Eq.~(146) because the latter specifies to modified theories without preferred frames, and theories with GW polarizations that travel at different speeds (e.g. Einstein-\ae{}ther theory) can have preferred frames.}:
\begin{widetext}
\begin{align}
    \tilde{h}(f) &= \tilde{h}_{GR} e^{i\beta u_2^b} + \left[ \alpha_+ F_+ + \alpha_\times F_\times + \alpha_b F_b + \alpha_L F_L + \alpha_X F_X + \alpha_Y F_Y \right] u_2^a e^{-i\Psi_{GR}^{(2)}} e^{i\beta u_2^b} \nonumber\\
    &\2\2 + \left[ \gamma_+ F_+ + \gamma_\times F_\times + \gamma_b F_b + \gamma_L F_L + \gamma_X F_X + \gamma_Y F_Y \right] u_1^c e^{-i\Psi_{GR}^{(1)}} e^{i\delta u_1^d}\,, 
    \label{eq:response-ppE}
\end{align}
\end{widetext}
where $\tilde{h}_{\text{GR}} = \mathcal{A}_{\text{GR}} u_2^{-7/2} e^{-i\Psi_{\text{GR}}^{(2)}}$ and $\{F_b, F_L, F_X, F_Y\}$ are the rest of the angular pattern functions as defined in~\cite{PWGravity2014}. Comparing this expression to Eq.~\eqref{eqn:ppE_all_harmonics}, we can see that $\{a, \beta, b, c, \delta, d\}$ are the same ppE parameters from before, but $\alpha\rightarrow\{\alpha_+, \alpha_\times, \alpha_b, \alpha_L, \alpha_X, \alpha_Y\}$ and $\gamma\rightarrow\{\gamma_+, \gamma_\times, \gamma_b, \gamma_L, \gamma_X, \gamma_Y\}$. Any possible inclination angle dependence has been absorbed within the $\alpha$ and $\gamma$ parameters. It is convenient to further condense the notation to 
\begin{align}
    \tilde{h}(f) &= \tilde{h}_{GR} e^{i\beta u_2^b} + \left[ \sum_N \alpha_N F_N \right] u_2^a e^{-i\Psi_{GR}^{(2)}} e^{i\beta u_2^b} \nonumber\\
    &\2\2 + \left[ \sum_N \gamma_N F_N \right] u_1^c e^{-i\Psi_{GR}^{(1)}} e^{i\delta u_1^d}
    \label{eqn:ppE_all_modes}
\end{align}
for $N \in \{+, \times, b, L, X, Y\}$.

In theories with different propagation speeds, the GW will pick up a phase factor that depends on this propagation speed. Since the propagation speeds of different polarizations can be different in the theories considered here, we need to include a phase factor for each polarization.\footnote{Note that this is different from Eq. (153) of~\cite{Chatziioannou:2012rf}, because that expression applied \textit{the same} phase change to all polarizations simultaneously, and that would not work for polarizations that have different speeds. }
Thus, if we are to generalize Eq.~\eqref{eqn:ppE_all_modes} to theories with multiple polarizations that propagate at different speeds, we have 
\begin{align}
    \tilde{h}(f) &= \tilde{h}_{GR} e^{i\beta u_2^b} \nonumber\\
    &\2\2+ \left[ \sum_N \alpha_N F_N e^{-2\pi if R (1 - c_N^{-1})}\right] u_2^a e^{-i\Psi_{GR}^{(2)}} e^{i\beta u_2^b} \nonumber\\
    &\2\2 + \left[ \sum_N \gamma_N F_N e^{-2\pi if R (1 - c_N^{-1})} \right] u_1^c e^{-i\Psi_{GR}^{(1)}} e^{i\delta u_1^d}\,, \label{eqn:ppEgeneralized}
\end{align}
where $c_N$ is the speed of each polarization, $\{c_+, c_\times, c_b, c_L, c_X, c_Y\}$. Note that not all of the speeds are distinct. For instance, when all scalar fields travel at $c_S$,  $c_B = c_L  = c_S$. Likewise, when vector fields travel at $c_V$,  $c_X = c_Y = c_V$. As previously discussed, $c_T$ has been well constrained to be equal to the speed of light by the GW170817 event, so we could also set $c_+ = c_\times = c$. Thus, if we set $c_T = c$, there are only two additional parameters introduced with this extension of the ppE framework, $c_S$ and $c_V$. The waveforms in Einstein-\ae{}ther theory derived in Eq.~(6.1) of~\cite{Zhang:2019iim} can be mapped to Eq.~\eqref{eqn:ppEgeneralized} above if we keep only the leading amplitude corrections for each polarization mode and absorb the inclination angle dependence into $\alpha_N$ and $\gamma_N$.\footnote{These expressions agree up to the sign of the phase change due to propagation speeds. However, the sign in Eq. (6.1) of~\cite{Zhang:2019iim} is inconsistent with the other expressions in that work, so it is likely a typo.}

\section{Null Streams}
\label{sec:null_streams}

As proposed in~\cite{Chatziioannou:2012rf}, one can construct null streams (which are ``null'' within GR) through suitable projections. These can be used to search for statistically significant deviations from noise and any detection would immediately signal a deviation from GR. Null streams can be used to search for additional polarizations and place constraints on those polarizations. Given $D$ detectors~\cite{Chatziioannou:2012rf},
\begin{equation}
    \begin{bmatrix}
        \tilde{d}_1 \\
        \tilde{d}_2 \\
        \vdots \\
        \tilde{d}_D \\
    \end{bmatrix} = 
    \begin{bmatrix}
        F^+_1 & F^\times_1 & F^X_1 & F^Y_1 & F^b_1 & F^L_1 \\
        F^+_2 & F^\times_2 & F^X_2 & F^Y_2 & F^b_2 & F^L_2 \\
        \vdots & \vdots & \vdots & \vdots & \vdots & \vdots \\
         F^+_D & F^\times_D & F^X_D & F^Y_D & F^b_D & F^L_D \\
    \end{bmatrix}
    \begin{bmatrix}
        \tilde{h}_+ \\
        \tilde{h}_\times \\
        \vdots \\
        \tilde{h}_L \\
    \end{bmatrix} +
    \begin{bmatrix}
        \tilde{n}_1 \\
        \tilde{n}_2 \\
        \vdots \\
        \tilde{n}_D \\
    \end{bmatrix}\,,
\end{equation}
where $\tilde{d}_a, F^\cdot_a$ and $\tilde{n}_a$ are the noise-weighted signal, the angular pattern functions, and the noise of the $a$th detector respectively, each normalized with respect to the power spectral density. In tensor notation, this becomes
\begin{equation}
    \tilde{d}^a = F^a_{\2j} \tilde{h}^j + \tilde{n}^a\,,
\end{equation}
where $a$ runs over the number of detectors and $j$ over the polarizations. In the above expressions, the $\tilde{h}_N$ polarizations must be generalized to include the $\alpha_N$ and $\gamma_N$ parameters of Eq.~\eqref{eq:response-ppE}. To create a data set with no component of a particular polarization, one must project the data set in a direction orthogonal to the angular pattern function for that particular polarization. Reference \cite{Chatziioannou:2012rf} does this for three detectors and for the stream that contains no tensor modes (the GR null stream): 
\begin{equation}
    \tilde{d}_{\text{GR, null}} = \frac{\epsilon^{cab} F^+_a F^\times_b}{|\delta^{ab}F^+_a F^\times_b|} \tilde{d}_c\,.
\end{equation}

All of the above equations are still applicable for theories in which the GW polarizations can travel at different speeds. The construction of the null stream is unaffected because the geometry of these polarizations is unchanged. However, in this case, there is an additional (implicit) dependence on the speed of propagation hidden in the $\tilde{h}_N$ terms. Clearly, different propagation speeds will result in different arrival times at the detector. Furthermore, the farther away the source is, the more propagation effects build up. Thus, even a small difference in propagation speeds could result in a large difference in arrival time if the signal had to travel a great distance to reach the detector. This potential for different arrival times poses challenges for placing constraints on additional polarizations via null streams. 

GW data analysis often considers a $32$ or $4096$ second window around the time of an event ($\pm 16$ or $\pm 2048$ seconds from the trigger)\footnote{Though one can work with the full data from the GW Open Science Center (gwosc.org), it is common to consider only the data surrounding a particular event and within one of these two time windows.}. Depending on the speed of propagation, additional polarizations could arrive either before or after this window and be missed entirely. Fortunately, observations of high energy cosmic rays rule out the possibility of gravitational Cherenkov radiation~\cite{Moore:2001bv, Kiyota:2015dla}. This forces the propagation speeds of GWs to be $c_{N} \geq c$, because if $c_N < c$ it would be possible for massive particles to travel faster than the GW, thereby producing Cherenkov radiation, which is not observed. As a result, any additional polarizations would have to arrive before or simultaneously with the tensor polarizations given that the propagation speed of the tensor mode has been constrained very stringently from GW170817.

As an example, let us consider a source of GWs that is 100 Mpc away. It would take GWs traveling at the speed of light $\approx 10^{16}$s to reach a detector from this distance. Meanwhile, polarizations traveling at speeds $c_N = c (1 + \mathcal{O}(10^{-13}))$\footnote{This argument assumes that $c_N > 1$ is allowed, as in Lorentz-violating theories that can have different propagation speeds, $c_N > 1$ does not violate causality.} will arrive 2048 seconds before the tensor polarizations, just outside the detector window (see Appendix~\ref{sec:appendixSpeeds} for details). Hence, even a very small difference in the speeds of the different polarizations could cause a null stream search to miss them entirely. Therefore, we caution that any null stream constraints on additional polarizations \textit{would only apply to polarizations that travel with speed $c_N = 1$}.

Considering different arrival times, is also possible that one could observe pure scalar or vector modes in gravitational wave data. Such polarizations cannot exist without the tensor polarizations of GR, yet they might arrive so far removed from the tensor polarizations as to be hard to associate with each other. For a source at 100 Mpc and a speed of $c_{N} = c(1 + \mathcal{O}(10^{-8}))$, the different polarizations could arrive 10 years apart, which is longer than we have so far been observing GW events. This is important to account for when considering searches for additional polarizations in GW data.

\section{Conclusions}
\label{sec:conclusions}
In this work, we have developed a straightforward method to compute the different GW polarizations directly from the metric perturbation, while allowing for the polarizations to travel with speeds $c_{\Scalar}, c_{\Vector}$, and $c_{\Tensor}$ that are different from the speed of light.
This result generalizes the method of~\cite{Chatziioannou:2012rf} to modified theories of gravity that excite vector or scalar degrees of freedom that propagate with different speeds. 
We present an expression that can be used with the metric perturbation, Eq.~\eqref{eqn:resultsDiffSpeeds}, and equivalent expressions that can be used with the trace-reversed metric perturbation, (Eqs.~\eqref{eqn:TRbmode}, ~\eqref{eqn:TRLmode},~\eqref{eqn:TRvectormode}), both of which are consistent with each other through the use of Eq.~\eqref{eqn:metricConversionRule}. 
We also showed that both of these expressions are consistent with previous work, \cite{Chatziioannou:2012rf}, in the limit that all of the speeds are equal to the speed of light: $c_{\Scalar,\Vector,\Tensor} \rightarrow c$.
We used this generic method to compute the GW polarizations in two specific modified theories of gravity: Einstein-\ae{}ther theory and khronometric gravity.
Our Einstein-\ae{}ther result matches that of previous work and serves as further confirmation of the validity of our method. 
The result we obtained for khronometric gravity corrects the result of~\cite{Hansen:2014ewa}, which used a method that is only valid when all degrees of freedom travel at the speed of light~\cite{Chatziioannou:2012rf}.
We further extended the work of~\cite{Chatziioannou:2012rf}, generalizing the ppE formalism to account for propagation effects and discussing the impact these propagation effects may have on null channel tests. We emphasize that even very small differences in speed could result in large differences in arrival times between modes, which would make it very challenging to conclusively rule out additional modes. This result may also motivate searches for pure scalar or vector modes.

Our results allow for the direct and simple evaluation of polarization modes using post-Newtonian and post-Minkowskian solutions to modified field equations in complex modified theories.
For instance, one might employ our formalism to construct polarization modes in f(R) gravity and Horndeski theory~\cite{Gong:2017bru}, or in linear massive gravity and generic curvature gravity~\cite{Tachinami:2021jnf}, since these theories contain massive modes with propagation speeds different from the speed of light.
Once the polarizations are obtained, further work to compute the detector response function and develop theory-specific waveform templates and tests is possible.
Alternatively, the ppE formalism that we have updated to include the effects of polarizations traveling at different speeds could be used in place of theory-specific templates.
As more detectors come online and more data becomes available, the polarization content of GWs will become increasingly important as a potential signature for new physics. The absence of such polarization content may greatly constrain a large class of modified theories of gravity, though only for certain propagation speeds.

\acknowledgments
K.S. would like to acknowledge that this material is based upon work supported by the National Science Foundation Graduate Research Fellowship Program under Grant No. DGE – 1746047.
N. Y. acknowledges support from NSF Grant No. AST 2009268 and No. PHY 2207650. 
K.Y. acknowledges support from NSF Grant PHY-2207349, PHY-2309066, a Sloan Foundation Research Fellowship and the Owens Family Foundation.

\appendix
\section{Polarization Modes with Different Speeds From the Trace-Reversed Metric Perturbation}
\label{sec:TRmetricPerturbation}
\renewcommand{\theequation}{\thesection.\arabic{equation}}

We here re-derive the GW polarizations from the \textit{trace-reversed} metric perturbation directly. 
Due to the different form of the linearized Riemann tensor when expressed in terms of the trace-reversed metric perturbation, we repeat elements of the calculation here to ensure the correctness of the derivation. 
After this is calculated, a more direct approach is again developed, using effective field theory techniques. 
We also establish a mapping between the scalar, vector, and tensor functions of the trace-reversed metric perturbation to those of the regular metric perturbation so that we can compare this result to those of Sec.~\ref{sec:metricPerturbation}.
From this calculation, it is clear that the trace-reversed version of Eq.~\eqref{eqn:resultsDiffSpeeds} could also be computed by applying the trace-reverse transformation of Eq.~\eqref{eqn:metricConversionRule}.

\subsection{Linearized Riemann Tensor Approach}
Let us begin by parameterizing the trace-reversed metric perturbation as 
\begin{align}
 \label{eqn:hbardecomp}
\h^{00} &= \frac{C}{c^4R}, \quad
\h^{0j} = \frac{D^j}{c^4R}, \quad
\h^{jk} = \frac{A^{jk}}{c^4R}.
\end{align}
Once again, the factor of $1/c^4 R$ will be absorbed into the functions $C, D^j$ and $A^{jk}$ from now on.
We once again use the fact that vectors and tensors can be irreducibly decomposed, so that 
\begin{align}
D^j &= \partial^j D + D^j_{\Tensor}\,,
\\
A^{jk} &= \frac{1}{3} \delta^{jk} A + \left(\partial^{jk} - \frac{1}{3}\delta^{jk} \nabla^2 \right)B + 2 \partial^{(j} A^{k)}_{\Tensor} + A^{jk}_{\TT} \,.
\end{align}
with $\partial_j A^j_{\Tensor} = 0$, $\partial_j A^{jk}_{\TT} = 0$, $\delta_{jk} A^{jk}_{\TT} = 0$, and $\partial_j D^j_{\Tensor} =0$.
As before, each of the fields above depends on $\N$ and retarded time $\tau$. 
Following the same steps as before, the irreducible decomposition of the trace-reversed metric perturbation can be rewritten as
\begin{subequations} \label{eqn:TRgravitationalPotentials}
\begin{align}
\h^{00} &= C(\tau_{\Scalar}, \N), \\
\h^{0j} &= \frac{1}{c_{\Scalar}} N^j D(\tau_{\Scalar}, \N) + D^j_{\Tensor} (\tau_{\Vector}, \N), \\
\h^{jk} &= \frac{\delta^{jk}}{3} A(\tau_{\Scalar}, \N) + \frac{1}{c_{\Scalar}^2} \left(N^j N^k - \frac{\delta^{jk}}{3} \right)B(\tau_{\Scalar}, \N) \nonumber \\
&\2\2+\frac{1}{c_{\Vector}} N^j A^k_{\Tensor}(\tau_{\Vector},\N) + \frac{1}{c_{\Vector}} N^k A^j_{\Tensor}(\tau_{\Vector}, N) \nonumber \\
&\2\2 + A^{jk}_{\TT}(\tau_{\Tensor}, \N).
\end{align}
\end{subequations}

Let us now consider the linearized Riemann tensor, which was already presented in Eq.~\eqref{eqn:Riemann}. 
Converting from the metric perturbation to the trace-reversed metric perturbation with Eq.~\eqref{eqn:metricConversionRule}, and following the index convention of \cite{PWGravity2014}, we have
\begin{multline}
R_{0j0k} = -\frac{1}{2} \left(\partial_{00} \h^{jk} - \frac{1}{2} \partial_{00} \h \delta_{jk} + \partial_{jk} \h^{00} \right.  \\
\left.\hspace{13mm}+ \frac{1}{2} \partial_{jk} \h + \partial_{0j} \h^{0k} + \partial_{0k} \h^{0j} \right).
\label{eqn:TRRiemann}
\end{multline} 
Now, inserting the irreducible decomposition of the trace-reversed metric perturbation into the linearized Riemann tensor and simplifying yields 
\begin{align}
R_{0j0k} &= -\frac{1}{2c^2} \partial^{\tt} \left[ \frac{\delta^{jk}}{3} A + \frac{1}{c_{\Scalar}^2} \left(N^j N^k - \frac{\delta^{jk}}{3}\right)B \right. \nonumber \\
&\2\2+\frac{N^j}{c_{\Vector}}A^k_{\Tensor} + \frac{N^k}{c_{\Vector}} A^j_{\Tensor} + A^{jk}_{\TT} + \frac{\delta^{jk}}{2} C \nonumber \\
&\2\2+ \frac{N^j N^k}{2} \frac{c^2}{c_{\Scalar}^2} C - \frac{\delta^{jk}}{2} A + \frac{N^j N^k}{2} \frac{c^2}{c_{\Scalar}^2} A \nonumber\\
& \left. \2\2-2N^j N^k \frac{c}{c_{\Scalar}^2} D - N^j \frac{c}{c_{\Vector}} D^k_{\Tensor} - N^k \frac{c}{c_{\Vector}} D^j_{\Tensor}\right]
\end{align}
This expression can be rearranged into the form 
\begin{multline}
R_{0j0k} = -\frac{1}{2c^2} \partial_{tt} \Big[(\delta^{jk} - N^j N^k) A_{\bb} + N^j N^k A_{\LL}  \\
\left. + 2N^{(j} A^{k)}_{\Vector} + A^{jk}_{\TT}\right]\,,
\end{multline}
where 
\begin{subequations}\label{eqn:TRresults}
\begin{align}
A_{\bb} &:= - \frac{1}{6}\left[A + \frac{2}{c_{\Scalar}^2} B - 3C\right], \label{eqn:TRbreathingMode}\\
A_{\LL} &:= \frac{1}{3} \left[ \left(\frac{3}{2}\frac{c^2}{c_{\Scalar}^2} - \frac{1}{2} \right)A + \frac{2}{c_{\Scalar}^2} B \right. \nonumber\\
&\hspace{10mm} \left. +\left(\frac{3}{2} \frac{c^2}{c_{\Scalar}^2} + \frac{3}{2}\right)C -6\frac{c}{c_{\Scalar}^2} D\right], \label{eqn:TRlongitudinalMode}\\
A^k_{\Vector} &:= \frac{1}{c_{\Vector}} \left[A^k_{\Tensor} - cD^k_{\Tensor}\right]. \label{eqn:TRvectorialMode}
\end{align}
\end{subequations}
As before, these are the scalar breathing mode, $A_{\bb}$, the scalar longitudinal mode, $A_{\LL}$, and the two vector modes, $A^k_{\Vector}$, expressed in terms of the decomposition of the trace-reversed metric perturbation. 

\subsection{Comparison with Previous Result} 
The result in Eq.~\eqref{eqn:TRresults}, derived from the trace-reversed metric perturbation, can be compared with that previously found in Eq.~\eqref{eqn:results}, which was derived from the metric perturbation. To do this, we need to map from the scalar, vector, and tensor functions $C, D^j, A^{jk}$ of the trace reversed metric perturbation to the scalar, vector, and tensor functions $\C, \D_j, \A_{jk}$ of the regular metric perturbation. Using the inverse of Eq.~\eqref{eqn:metricConversionRule}, we note that 
\begin{subequations}
\begin{align}
\h^{00} &=  \frac{1}{2} \left(h_{00} + h_{kk} \right)\,, \\
\h^{0j} &= - h_{0j}\,, \\
\h^{jk} &= h_{jk} + \frac{1}{2} \delta_{jk} \left( h_{00}- h_{kk} \right)\,,
\end{align}
\end{subequations}
and then, by Eqs.~\eqref{eqn:hdecomp} and \eqref{eqn:hbardecomp}, we have that 
\begin{subequations}
\begin{align}
C &= \frac{1}{2} \left(\C + \A\right) \\
D^j &= - \D_j \\
A^{jk} &= \frac{1}{3} \delta_{jk} \left( \frac{3}{2} \C - \frac{1}{2} \A \right) + \left(\partial_{jk} - \frac{1}{3} \delta_{jk} \nabla^2 \right) \B \nonumber \\
&\2\2 + \partial_j \A^T_k + \partial_k \A^T_j + \A^{\TT}_{jk} 
\end{align} 
\end{subequations}
Recalling the decompositions we have already established for $D^j$ and $A^{jk}$ and comparing like terms, 
we arrive at
\begin{align}
\begin{aligned} \label{eqn:mapping}
C &= \frac{1}{2} \left(\C + \A \right), \quad
D = - \D, \quad
D^j_{\Tensor} = -\D^T_j,
\\
A &= \frac{1}{2} \left(3\C - \A\right), \quad
B = \B, \quad
A^j_{\Tensor} = \A^T_j, \quad
A^{jk}_{\TT} = \A^{\TT}_{jk}\,.
\end{aligned}
\end{align}
Inserting these expressions into Eq.~\eqref{eqn:TRresults} returns exactly Eq.~\eqref{eqn:results}. 

\subsection{An Effective Field Theory Approach}
Let us now repeat the calculation we previously carried out using effective field theory, but this time apply it to the trace-reversed metric perturbation. 
We thus postulate the ansatz
\begin{equation}
A_{\bb} = \bar{a}_1 \h^{00} + \bar{a}_2 N_j \h^{0j} + \bar{a}_3 N_j N_k \h^{jk} + \bar{a}_4 \delta_{jk} \h^{jk}\,,
\end{equation}
which we then insert in Eq.~\eqref{eqn:TRgravitationalPotentials} and simplify to obtain 
\begin{equation}
A_{\bb} = \bar{a}_1 C + \frac{\bar{a}_2}{c_{\Scalar}} D + \bar{a}_3 \left(\frac{1}{3} A + \frac{2}{3c_{\Scalar}^2} B\right) + \bar{a}_4 A.
\end{equation}
Comparing with Eq.~\eqref{eqn:breathingMode} and matching coefficients, we find
\begin{equation}
\bar{a}_1 = \frac{1}{2},\2 \bar{a}_2 = 0,\2 \bar{a}_3 = -\frac{1}{2},\2 \bar{a}_4 = 0\,,
\end{equation}
and therefore,
\begin{equation}
A_{\bb} = \frac{1}{2} \left(\h^{00} - N_j N_k \h^{jk}\right). \label{eqn:TRbmode}
\end{equation}
Following the same procedure for the scalar longitudinal mode, we find
\begin{align}
A_{\LL} &= \frac{1}{2} \left(\frac{c^2}{c_{\Scalar}^2} + 1\right)\h^{00} -2\frac{c}{c_{\Scalar}} N_j \h^{0j} 
\nonumber \\
&+ \left[N_j N_k + \frac{1}{2} \left(\frac{c^2}{c_{\Scalar}^2} - 1\right)\delta_{jk}\right] \h^{jk}.\label{eqn:TRLmode}
\end{align}

Similarly, to derive the vector modes, we consider the ansatz 
\begin{equation}
A^k_{\Vector} = P^k_j \left(\bar{a}_5\h^{0j} + \bar{a}_6 N_i \h^{ij}\right),
\end{equation}
with some constants $\bar{a}_5$ and $\bar{a}_6$.
Inserting in Eq.~\eqref{eqn:gravitationalPotentials} and simplifying, we find 
\begin{equation}
A^k_{\Vector} = \bar{a}_5 D^k_{\Tensor} + \bar{a}_6 \frac{1}{c_{\Vector}} A^k_{\Tensor}\,,
\end{equation}
and matching coefficients with Eq.\eqref{eqn:vectorialMode} gives
\begin{align}
\bar{a}_5 = -\frac{c}{c_{\Vector}},\2 \bar{a}_6 = 1\,,
\end{align}
so that then 
\begin{equation}
A^k_{\Vector} = P^k_j\left(N_i\h^{ij} - \frac{c}{c_{\Vector}} \h^{0j}\right). 
\label{eqn:TRvectormode}
\end{equation}
These expressions, calculated directly from the trace-reversed metric perturbation, are consistent with what was found from the linearized Riemann tensor calculation (Eq.~\eqref{eqn:TRresults}). 
Although these expression may look different from Eq.~\eqref{eqn:resultsDiffSpeeds}, a quick calculation with Eq.~\eqref{eqn:metricConversionRule} reveals that the two expressions are indeed, consistent.

\subsection{Comparison to previous results} 
As a test of our expressions in the previous section, we can take the limit where all modes travel at the speed of light, and compare to Chatziioannou \emph{et. al}~\cite{Chatziioannou:2012rf}. If $c_{\Scalar,\Vector,\Tensor} \rightarrow c$, then the previous expressions (Eqs.~\eqref{eqn:TRbmode}, ~\eqref{eqn:TRLmode},~\eqref{eqn:TRvectormode}) reduce to
\begin{subequations}
\begin{align}
A_{\bb} &= \frac{1}{2} \left(\h^{00} - N_j N_k \h^{jk} \right), \label{eqn:TRbreathingModeSameSpeeds}\\
A_{\LL} &= \h^{00} - 2N_j \h^{0j} + N_j N_k \h^{jk}, \label{eqn:TRlongitudinalModeSameSpeeds} \\
A^k_{\Vector} &= P^k_j \left(N_i \h^{ij} - \h^{0j} \right). \label{eqn:TRvectorialModeSameSpeeds}
\end{align}
\end{subequations}
This corresponds to Eqs.~\eqref{eqn:finalResultSameSpeeds}, and hence, it matches what was found by Chatziioannou \emph{et. al} in Eqs.~(9a)--(9c) of~\cite{Chatziioannou:2012rf} (up to the overall minus sign on $A_{\bb}$).

\section{Polarization speeds and detector arrival times}
\label{sec:appendixSpeeds}
For a given polarization of the GW, 
\begin{equation}
    t_N = \frac{R}{c_N}\,,
\end{equation}
where $t_N$ is the arrival time of the $N^{th}$ mode, $R$ is the distance to the source, and $c_N$ is the speed of that polarization. For two different polarizations, $A$ and $B$, the difference in arrival times is 
\begin{equation}
    \Delta t_{A,B} = t_{A} - t_{B} = \frac{R}{c_{A}} - \frac{R}{c_{B}}\,.
\end{equation}
From this it is a simple matter to compute what speed of GW would be necessary for a particular difference in arrival times:
\begin{align}
    c_{B} &= \left(\frac{1}{c_{A}} - \frac{\Delta t_{A,B}}{R} \right)^{-1}\,.
\end{align}

Consider a source with $R = 100$ Mpc $\approx 10^{24}$ m. The tensor polarizations travel at the speed of light, $c \approx 3 \times 10^8$ m/s. For a difference in arrival times between one of the tensor modes and a different polarization, of at least $\Delta t_{+, N} = 2048$ seconds, we would need $c_{N} \gtrsim c(1 + \mathcal{O}(10^{-13}))$. This would be a great enough difference for the additional polarizations to not appear in the standard detector window. For a difference in arrival times of ten years ($\approx 3 \times 10^8$ seconds), which is longer than we have been observing GW events, we would need $c_N \approx c(1 + \mathcal{O}(10^{-8}))$.

\bibliography{main, notInspire}

\begin{thebibliography}{40}%
\makeatletter
\providecommand \@ifxundefined [1]{%
 \@ifx{#1\undefined}
}%
\providecommand \@ifnum [1]{%
 \ifnum #1\expandafter \@firstoftwo
 \else \expandafter \@secondoftwo
 \fi
}%
\providecommand \@ifx [1]{%
 \ifx #1\expandafter \@firstoftwo
 \else \expandafter \@secondoftwo
 \fi
}%
\providecommand \natexlab [1]{#1}%
\providecommand \enquote  [1]{``#1''}%
\providecommand \bibnamefont  [1]{#1}%
\providecommand \bibfnamefont [1]{#1}%
\providecommand \citenamefont [1]{#1}%
\providecommand \href@noop [0]{\@secondoftwo}%
\providecommand \href [0]{\begingroup \@sanitize@url \@href}%
\providecommand \@href[1]{\@@startlink{#1}\@@href}%
\providecommand \@@href[1]{\endgroup#1\@@endlink}%
\providecommand \@sanitize@url [0]{\catcode `\\12\catcode `\$12\catcode
  `\&12\catcode `\#12\catcode `\^12\catcode `\_12\catcode `\%12\relax}%
\providecommand \@@startlink[1]{}%
\providecommand \@@endlink[0]{}%
\providecommand \url  [0]{\begingroup\@sanitize@url \@url }%
\providecommand \@url [1]{\endgroup\@href {#1}{\urlprefix }}%
\providecommand \urlprefix  [0]{URL }%
\providecommand \Eprint [0]{\href }%
\providecommand \doibase [0]{https://doi.org/}%
\providecommand \selectlanguage [0]{\@gobble}%
\providecommand \bibinfo  [0]{\@secondoftwo}%
\providecommand \bibfield  [0]{\@secondoftwo}%
\providecommand \translation [1]{[#1]}%
\providecommand \BibitemOpen [0]{}%
\providecommand \bibitemStop [0]{}%
\providecommand \bibitemNoStop [0]{.\EOS\space}%
\providecommand \EOS [0]{\spacefactor3000\relax}%
\providecommand \BibitemShut  [1]{\csname bibitem#1\endcsname}%
\let\auto@bib@innerbib\@empty
\bibitem [{\citenamefont {Will}(2014)}]{Will:2014kxa}%
  \BibitemOpen
  \bibfield  {author} {\bibinfo {author} {\bibfnamefont {C.~M.}\ \bibnamefont
  {Will}},\ }\bibfield  {title} {\bibinfo {title} {{The Confrontation between
  General Relativity and Experiment}},\ }\href
  {https://doi.org/10.12942/lrr-2014-4} {\bibfield  {journal} {\bibinfo
  {journal} {Living Rev. Rel.}\ }\textbf {\bibinfo {volume} {17}},\ \bibinfo
  {pages} {4} (\bibinfo {year} {2014})},\ \Eprint
  {https://arxiv.org/abs/1403.7377} {arXiv:1403.7377 [gr-qc]} \BibitemShut
  {NoStop}%
\bibitem [{\citenamefont {Clifton}\ \emph {et~al.}(2012)\citenamefont
  {Clifton}, \citenamefont {Ferreira}, \citenamefont {Padilla},\ and\
  \citenamefont {Skordis}}]{Clifton:2011jh}%
  \BibitemOpen
  \bibfield  {author} {\bibinfo {author} {\bibfnamefont {T.}~\bibnamefont
  {Clifton}}, \bibinfo {author} {\bibfnamefont {P.~G.}\ \bibnamefont
  {Ferreira}}, \bibinfo {author} {\bibfnamefont {A.}~\bibnamefont {Padilla}},\
  and\ \bibinfo {author} {\bibfnamefont {C.}~\bibnamefont {Skordis}},\
  }\bibfield  {title} {\bibinfo {title} {{Modified Gravity and Cosmology}},\
  }\href {https://doi.org/10.1016/j.physrep.2012.01.001} {\bibfield  {journal}
  {\bibinfo  {journal} {Phys. Rept.}\ }\textbf {\bibinfo {volume} {513}},\
  \bibinfo {pages} {1} (\bibinfo {year} {2012})},\ \Eprint
  {https://arxiv.org/abs/1106.2476} {arXiv:1106.2476 [astro-ph.CO]}
  \BibitemShut {NoStop}%
\bibitem [{\citenamefont {Abbott}\ \emph {et~al.}(2016)\citenamefont {Abbott}
  \emph {et~al.}}]{LIGOScientific:2016lio}%
  \BibitemOpen
  \bibfield  {author} {\bibinfo {author} {\bibfnamefont {B.~P.}\ \bibnamefont
  {Abbott}} \emph {et~al.} (\bibinfo {collaboration} {LIGO Scientific,
  Virgo}),\ }\bibfield  {title} {\bibinfo {title} {{Tests of general relativity
  with GW150914}},\ }\href {https://doi.org/10.1103/PhysRevLett.116.221101}
  {\bibfield  {journal} {\bibinfo  {journal} {Phys. Rev. Lett.}\ }\textbf
  {\bibinfo {volume} {116}},\ \bibinfo {pages} {221101} (\bibinfo {year}
  {2016})},\ \bibinfo {note} {[Erratum: Phys.Rev.Lett. 121, 129902 (2018)]},\
  \Eprint {https://arxiv.org/abs/1602.03841} {arXiv:1602.03841 [gr-qc]}
  \BibitemShut {NoStop}%
\bibitem [{\citenamefont {Yunes}\ \emph {et~al.}(2016)\citenamefont {Yunes},
  \citenamefont {Yagi},\ and\ \citenamefont {Pretorius}}]{Yunes:2016jcc}%
  \BibitemOpen
  \bibfield  {author} {\bibinfo {author} {\bibfnamefont {N.}~\bibnamefont
  {Yunes}}, \bibinfo {author} {\bibfnamefont {K.}~\bibnamefont {Yagi}},\ and\
  \bibinfo {author} {\bibfnamefont {F.}~\bibnamefont {Pretorius}},\ }\bibfield
  {title} {\bibinfo {title} {{Theoretical Physics Implications of the Binary
  Black-Hole Mergers GW150914 and GW151226}},\ }\href
  {https://doi.org/10.1103/PhysRevD.94.084002} {\bibfield  {journal} {\bibinfo
  {journal} {Phys. Rev. D}\ }\textbf {\bibinfo {volume} {94}},\ \bibinfo
  {pages} {084002} (\bibinfo {year} {2016})},\ \Eprint
  {https://arxiv.org/abs/1603.08955} {arXiv:1603.08955 [gr-qc]} \BibitemShut
  {NoStop}%
\bibitem [{\citenamefont {Abbott}\ \emph
  {et~al.}(2019{\natexlab{a}})\citenamefont {Abbott} \emph
  {et~al.}}]{LIGOScientific:2018dkp}%
  \BibitemOpen
  \bibfield  {author} {\bibinfo {author} {\bibfnamefont {B.~P.}\ \bibnamefont
  {Abbott}} \emph {et~al.} (\bibinfo {collaboration} {LIGO Scientific,
  Virgo}),\ }\bibfield  {title} {\bibinfo {title} {{Tests of General Relativity
  with GW170817}},\ }\href {https://doi.org/10.1103/PhysRevLett.123.011102}
  {\bibfield  {journal} {\bibinfo  {journal} {Phys. Rev. Lett.}\ }\textbf
  {\bibinfo {volume} {123}},\ \bibinfo {pages} {011102} (\bibinfo {year}
  {2019}{\natexlab{a}})},\ \Eprint {https://arxiv.org/abs/1811.00364}
  {arXiv:1811.00364 [gr-qc]} \BibitemShut {NoStop}%
\bibitem [{\citenamefont {Berti}\ \emph
  {et~al.}(2018{\natexlab{a}})\citenamefont {Berti}, \citenamefont {Yagi},\
  and\ \citenamefont {Yunes}}]{Berti:2018cxi}%
  \BibitemOpen
  \bibfield  {author} {\bibinfo {author} {\bibfnamefont {E.}~\bibnamefont
  {Berti}}, \bibinfo {author} {\bibfnamefont {K.}~\bibnamefont {Yagi}},\ and\
  \bibinfo {author} {\bibfnamefont {N.}~\bibnamefont {Yunes}},\ }\bibfield
  {title} {\bibinfo {title} {{Extreme Gravity Tests with Gravitational Waves
  from Compact Binary Coalescences: (I) Inspiral-Merger}},\ }\href
  {https://doi.org/10.1007/s10714-018-2362-8} {\bibfield  {journal} {\bibinfo
  {journal} {Gen. Rel. Grav.}\ }\textbf {\bibinfo {volume} {50}},\ \bibinfo
  {pages} {46} (\bibinfo {year} {2018}{\natexlab{a}})},\ \Eprint
  {https://arxiv.org/abs/1801.03208} {arXiv:1801.03208 [gr-qc]} \BibitemShut
  {NoStop}%
\bibitem [{\citenamefont {Berti}\ \emph
  {et~al.}(2018{\natexlab{b}})\citenamefont {Berti}, \citenamefont {Yagi},
  \citenamefont {Yang},\ and\ \citenamefont {Yunes}}]{Berti:2018vdi}%
  \BibitemOpen
  \bibfield  {author} {\bibinfo {author} {\bibfnamefont {E.}~\bibnamefont
  {Berti}}, \bibinfo {author} {\bibfnamefont {K.}~\bibnamefont {Yagi}},
  \bibinfo {author} {\bibfnamefont {H.}~\bibnamefont {Yang}},\ and\ \bibinfo
  {author} {\bibfnamefont {N.}~\bibnamefont {Yunes}},\ }\bibfield  {title}
  {\bibinfo {title} {{Extreme Gravity Tests with Gravitational Waves from
  Compact Binary Coalescences: (II) Ringdown}},\ }\href
  {https://doi.org/10.1007/s10714-018-2372-6} {\bibfield  {journal} {\bibinfo
  {journal} {Gen. Rel. Grav.}\ }\textbf {\bibinfo {volume} {50}},\ \bibinfo
  {pages} {49} (\bibinfo {year} {2018}{\natexlab{b}})},\ \Eprint
  {https://arxiv.org/abs/1801.03587} {arXiv:1801.03587 [gr-qc]} \BibitemShut
  {NoStop}%
\bibitem [{\citenamefont {Abbott}\ \emph
  {et~al.}(2019{\natexlab{b}})\citenamefont {Abbott} \emph
  {et~al.}}]{LIGOScientific:2019fpa}%
  \BibitemOpen
  \bibfield  {author} {\bibinfo {author} {\bibfnamefont {B.~P.}\ \bibnamefont
  {Abbott}} \emph {et~al.} (\bibinfo {collaboration} {LIGO Scientific,
  Virgo}),\ }\bibfield  {title} {\bibinfo {title} {{Tests of General Relativity
  with the Binary Black Hole Signals from the LIGO-Virgo Catalog GWTC-1}},\
  }\href {https://doi.org/10.1103/PhysRevD.100.104036} {\bibfield  {journal}
  {\bibinfo  {journal} {Phys. Rev. D}\ }\textbf {\bibinfo {volume} {100}},\
  \bibinfo {pages} {104036} (\bibinfo {year} {2019}{\natexlab{b}})},\ \Eprint
  {https://arxiv.org/abs/1903.04467} {arXiv:1903.04467 [gr-qc]} \BibitemShut
  {NoStop}%
\bibitem [{\citenamefont {Abbott}\ \emph
  {et~al.}(2021{\natexlab{a}})\citenamefont {Abbott} \emph
  {et~al.}}]{LIGOScientific:2020tif}%
  \BibitemOpen
  \bibfield  {author} {\bibinfo {author} {\bibfnamefont {R.}~\bibnamefont
  {Abbott}} \emph {et~al.} (\bibinfo {collaboration} {LIGO Scientific,
  Virgo}),\ }\bibfield  {title} {\bibinfo {title} {{Tests of general relativity
  with binary black holes from the second LIGO-Virgo gravitational-wave
  transient catalog}},\ }\href {https://doi.org/10.1103/PhysRevD.103.122002}
  {\bibfield  {journal} {\bibinfo  {journal} {Phys. Rev. D}\ }\textbf {\bibinfo
  {volume} {103}},\ \bibinfo {pages} {122002} (\bibinfo {year}
  {2021}{\natexlab{a}})},\ \Eprint {https://arxiv.org/abs/2010.14529}
  {arXiv:2010.14529 [gr-qc]} \BibitemShut {NoStop}%
\bibitem [{\citenamefont {Abbott}\ \emph
  {et~al.}(2021{\natexlab{b}})\citenamefont {Abbott} \emph
  {et~al.}}]{LIGOScientific:2021sio}%
  \BibitemOpen
  \bibfield  {author} {\bibinfo {author} {\bibfnamefont {R.}~\bibnamefont
  {Abbott}} \emph {et~al.} (\bibinfo {collaboration} {LIGO Scientific, VIRGO,
  KAGRA}),\ }\bibfield  {title} {\bibinfo {title} {{Tests of General Relativity
  with GWTC-3}},\ }\href@noop {} {\  (\bibinfo {year} {2021}{\natexlab{b}})},\
  \Eprint {https://arxiv.org/abs/2112.06861} {arXiv:2112.06861 [gr-qc]}
  \BibitemShut {NoStop}%
\bibitem [{\citenamefont {Abbott}\ \emph
  {et~al.}(2021{\natexlab{c}})\citenamefont {Abbott} \emph
  {et~al.}}]{LIGOScientific:2021djp}%
  \BibitemOpen
  \bibfield  {author} {\bibinfo {author} {\bibfnamefont {R.}~\bibnamefont
  {Abbott}} \emph {et~al.} (\bibinfo {collaboration} {LIGO Scientific, VIRGO,
  KAGRA}),\ }\bibfield  {title} {\bibinfo {title} {{GWTC-3: Compact Binary
  Coalescences Observed by LIGO and Virgo During the Second Part of the Third
  Observing Run}},\ }\href@noop {} {\  (\bibinfo {year}
  {2021}{\natexlab{c}})},\ \Eprint {https://arxiv.org/abs/2111.03606}
  {arXiv:2111.03606 [gr-qc]} \BibitemShut {NoStop}%
\bibitem [{\citenamefont {Abbott}\ \emph {et~al.}(2018)\citenamefont {Abbott}
  \emph {et~al.}}]{KAGRA:2013rdx}%
  \BibitemOpen
  \bibfield  {author} {\bibinfo {author} {\bibfnamefont {B.~P.}\ \bibnamefont
  {Abbott}} \emph {et~al.} (\bibinfo {collaboration} {KAGRA, LIGO Scientific,
  Virgo, VIRGO}),\ }\bibfield  {title} {\bibinfo {title} {{Prospects for
  observing and localizing gravitational-wave transients with Advanced LIGO,
  Advanced Virgo and KAGRA}},\ }\href
  {https://doi.org/10.1007/s41114-020-00026-9} {\bibfield  {journal} {\bibinfo
  {journal} {Living Rev. Rel.}\ }\textbf {\bibinfo {volume} {21}},\ \bibinfo
  {pages} {3} (\bibinfo {year} {2018})},\ \Eprint
  {https://arxiv.org/abs/1304.0670} {arXiv:1304.0670 [gr-qc]} \BibitemShut
  {NoStop}%
\bibitem [{\citenamefont {Jacobson}(2007)}]{Jacobson:2007veq}%
  \BibitemOpen
  \bibfield  {author} {\bibinfo {author} {\bibfnamefont {T.}~\bibnamefont
  {Jacobson}},\ }\bibfield  {title} {\bibinfo {title} {{Einstein-aether
  gravity: A Status report}},\ }\href {https://doi.org/10.22323/1.043.0020}
  {\bibfield  {journal} {\bibinfo  {journal} {PoS}\ }\textbf {\bibinfo {volume}
  {QG-PH}},\ \bibinfo {pages} {020} (\bibinfo {year} {2007})},\ \Eprint
  {https://arxiv.org/abs/0801.1547} {arXiv:0801.1547 [gr-qc]} \BibitemShut
  {NoStop}%
\bibitem [{\citenamefont {Mattingly}(2005)}]{Mattingly:2005re}%
  \BibitemOpen
  \bibfield  {author} {\bibinfo {author} {\bibfnamefont {D.}~\bibnamefont
  {Mattingly}},\ }\bibfield  {title} {\bibinfo {title} {{Modern tests of
  Lorentz invariance}},\ }\href {https://doi.org/10.12942/lrr-2005-5}
  {\bibfield  {journal} {\bibinfo  {journal} {Living Rev. Rel.}\ }\textbf
  {\bibinfo {volume} {8}},\ \bibinfo {pages} {5} (\bibinfo {year} {2005})},\
  \Eprint {https://arxiv.org/abs/gr-qc/0502097} {arXiv:gr-qc/0502097}
  \BibitemShut {NoStop}%
\bibitem [{\citenamefont {Sarbach}\ \emph {et~al.}(2019)\citenamefont
  {Sarbach}, \citenamefont {Barausse},\ and\ \citenamefont
  {Preciado-L\'opez}}]{Sarbach:2019yso}%
  \BibitemOpen
  \bibfield  {author} {\bibinfo {author} {\bibfnamefont {O.}~\bibnamefont
  {Sarbach}}, \bibinfo {author} {\bibfnamefont {E.}~\bibnamefont {Barausse}},\
  and\ \bibinfo {author} {\bibfnamefont {J.~A.}\ \bibnamefont
  {Preciado-L\'opez}},\ }\bibfield  {title} {\bibinfo {title} {{Well-posed
  Cauchy formulation for Einstein-\ae{}ther theory}},\ }\href
  {https://doi.org/10.1088/1361-6382/ab2e13} {\bibfield  {journal} {\bibinfo
  {journal} {Class. Quant. Grav.}\ }\textbf {\bibinfo {volume} {36}},\ \bibinfo
  {pages} {165007} (\bibinfo {year} {2019})},\ \Eprint
  {https://arxiv.org/abs/1902.05130} {arXiv:1902.05130 [gr-qc]} \BibitemShut
  {NoStop}%
\bibitem [{\citenamefont {Oost}\ \emph {et~al.}(2018)\citenamefont {Oost},
  \citenamefont {Mukohyama},\ and\ \citenamefont {Wang}}]{Oost:2018tcv}%
  \BibitemOpen
  \bibfield  {author} {\bibinfo {author} {\bibfnamefont {J.}~\bibnamefont
  {Oost}}, \bibinfo {author} {\bibfnamefont {S.}~\bibnamefont {Mukohyama}},\
  and\ \bibinfo {author} {\bibfnamefont {A.}~\bibnamefont {Wang}},\ }\bibfield
  {title} {\bibinfo {title} {{Constraints on Einstein-aether theory after
  GW170817}},\ }\href {https://doi.org/10.1103/PhysRevD.97.124023} {\bibfield
  {journal} {\bibinfo  {journal} {Phys. Rev. D}\ }\textbf {\bibinfo {volume}
  {97}},\ \bibinfo {pages} {124023} (\bibinfo {year} {2018})},\ \Eprint
  {https://arxiv.org/abs/1802.04303} {arXiv:1802.04303 [gr-qc]} \BibitemShut
  {NoStop}%
\bibitem [{\citenamefont {Schumacher}\ \emph {et~al.}(2023)\citenamefont
  {Schumacher}, \citenamefont {Perkins}, \citenamefont {Shaw}, \citenamefont
  {Yagi},\ and\ \citenamefont {Yunes}}]{Schumacher:2023cxh}%
  \BibitemOpen
  \bibfield  {author} {\bibinfo {author} {\bibfnamefont {K.}~\bibnamefont
  {Schumacher}}, \bibinfo {author} {\bibfnamefont {S.~E.}\ \bibnamefont
  {Perkins}}, \bibinfo {author} {\bibfnamefont {A.}~\bibnamefont {Shaw}},
  \bibinfo {author} {\bibfnamefont {K.}~\bibnamefont {Yagi}},\ and\ \bibinfo
  {author} {\bibfnamefont {N.}~\bibnamefont {Yunes}},\ }\bibfield  {title}
  {\bibinfo {title} {{Gravitational wave constraints on Einstein-\ae{}ther
  theory with LIGO/Virgo data}},\ }\href@noop {} {\  (\bibinfo {year}
  {2023})},\ \Eprint {https://arxiv.org/abs/2304.06801} {arXiv:2304.06801
  [gr-qc]} \BibitemShut {NoStop}%
\bibitem [{\citenamefont {Chatziioannou}\ \emph {et~al.}(2012)\citenamefont
  {Chatziioannou}, \citenamefont {Yunes},\ and\ \citenamefont
  {Cornish}}]{Chatziioannou:2012rf}%
  \BibitemOpen
  \bibfield  {author} {\bibinfo {author} {\bibfnamefont {K.}~\bibnamefont
  {Chatziioannou}}, \bibinfo {author} {\bibfnamefont {N.}~\bibnamefont
  {Yunes}},\ and\ \bibinfo {author} {\bibfnamefont {N.}~\bibnamefont
  {Cornish}},\ }\bibfield  {title} {\bibinfo {title} {{Model-Independent Test
  of General Relativity: An Extended post-Einsteinian Framework with Complete
  Polarization Content}},\ }\href {https://doi.org/10.1103/PhysRevD.86.022004}
  {\bibfield  {journal} {\bibinfo  {journal} {Phys. Rev. D}\ }\textbf {\bibinfo
  {volume} {86}},\ \bibinfo {pages} {022004} (\bibinfo {year} {2012})},\
  \bibinfo {note} {[Erratum: Phys.Rev.D 95, 129901 (2017)]},\ \Eprint
  {https://arxiv.org/abs/1204.2585} {arXiv:1204.2585 [gr-qc]} \BibitemShut
  {NoStop}%
\bibitem [{\citenamefont {Abbott}\ \emph {et~al.}(2017)\citenamefont {Abbott}
  \emph {et~al.}}]{LIGOScientific:2017zic}%
  \BibitemOpen
  \bibfield  {author} {\bibinfo {author} {\bibfnamefont {B.~P.}\ \bibnamefont
  {Abbott}} \emph {et~al.} (\bibinfo {collaboration} {LIGO Scientific, Virgo,
  Fermi-GBM, INTEGRAL}),\ }\bibfield  {title} {\bibinfo {title} {{Gravitational
  Waves and Gamma-rays from a Binary Neutron Star Merger: GW170817 and GRB
  170817A}},\ }\href {https://doi.org/10.3847/2041-8213/aa920c} {\bibfield
  {journal} {\bibinfo  {journal} {Astrophys. J. Lett.}\ }\textbf {\bibinfo
  {volume} {848}},\ \bibinfo {pages} {L13} (\bibinfo {year} {2017})},\ \Eprint
  {https://arxiv.org/abs/1710.05834} {arXiv:1710.05834 [astro-ph.HE]}
  \BibitemShut {NoStop}%
\bibitem [{\citenamefont {Poisson}\ and\ \citenamefont
  {Will}(2014)}]{PWGravity2014}%
  \BibitemOpen
  \bibfield  {author} {\bibinfo {author} {\bibfnamefont {E.}~\bibnamefont
  {Poisson}}\ and\ \bibinfo {author} {\bibfnamefont {C.~M.}\ \bibnamefont
  {Will}},\ }\href@noop {} {\emph {\bibinfo {title} {Gravity: {N}ewtonian,
  {P}ost-{N}ewtonian, {R}elativistic}}}\ (\bibinfo  {publisher} {Cambridge
  University Press},\ \bibinfo {year} {2014})\ pp.\ \bibinfo {pages} {540--546
  and 733--738}\BibitemShut {NoStop}%
\bibitem [{\citenamefont {Zhang}\ \emph {et~al.}(2020)\citenamefont {Zhang},
  \citenamefont {Zhao}, \citenamefont {Wang}, \citenamefont {Wang},
  \citenamefont {Yagi}, \citenamefont {Yunes}, \citenamefont {Zhao},\ and\
  \citenamefont {Zhu}}]{Zhang:2019iim}%
  \BibitemOpen
  \bibfield  {author} {\bibinfo {author} {\bibfnamefont {C.}~\bibnamefont
  {Zhang}}, \bibinfo {author} {\bibfnamefont {X.}~\bibnamefont {Zhao}},
  \bibinfo {author} {\bibfnamefont {A.}~\bibnamefont {Wang}}, \bibinfo {author}
  {\bibfnamefont {B.}~\bibnamefont {Wang}}, \bibinfo {author} {\bibfnamefont
  {K.}~\bibnamefont {Yagi}}, \bibinfo {author} {\bibfnamefont {N.}~\bibnamefont
  {Yunes}}, \bibinfo {author} {\bibfnamefont {W.}~\bibnamefont {Zhao}},\ and\
  \bibinfo {author} {\bibfnamefont {T.}~\bibnamefont {Zhu}},\ }\bibfield
  {title} {\bibinfo {title} {{Gravitational waves from the quasicircular
  inspiral of compact binaries in Einstein-aether theory}},\ }\href
  {https://doi.org/10.1103/PhysRevD.104.069905} {\bibfield  {journal} {\bibinfo
   {journal} {Phys. Rev. D}\ }\textbf {\bibinfo {volume} {101}},\ \bibinfo
  {pages} {044002} (\bibinfo {year} {2020})},\ \bibinfo {note} {[Erratum:
  Phys.Rev.D 104, 069905 (2021)]},\ \Eprint {https://arxiv.org/abs/1911.10278}
  {arXiv:1911.10278 [gr-qc]} \BibitemShut {NoStop}%
\bibitem [{\citenamefont {Horava}(2009)}]{Horava:2009uw}%
  \BibitemOpen
  \bibfield  {author} {\bibinfo {author} {\bibfnamefont {P.}~\bibnamefont
  {Horava}},\ }\bibfield  {title} {\bibinfo {title} {{Quantum Gravity at a
  Lifshitz Point}},\ }\href {https://doi.org/10.1103/PhysRevD.79.084008}
  {\bibfield  {journal} {\bibinfo  {journal} {Phys. Rev. D}\ }\textbf {\bibinfo
  {volume} {79}},\ \bibinfo {pages} {084008} (\bibinfo {year} {2009})},\
  \Eprint {https://arxiv.org/abs/0901.3775} {arXiv:0901.3775 [hep-th]}
  \BibitemShut {NoStop}%
\bibitem [{\citenamefont {Yunes}\ and\ \citenamefont
  {Pretorius}(2009)}]{Yunes:2009ke}%
  \BibitemOpen
  \bibfield  {author} {\bibinfo {author} {\bibfnamefont {N.}~\bibnamefont
  {Yunes}}\ and\ \bibinfo {author} {\bibfnamefont {F.}~\bibnamefont
  {Pretorius}},\ }\bibfield  {title} {\bibinfo {title} {{Fundamental
  Theoretical Bias in Gravitational Wave Astrophysics and the Parameterized
  Post-Einsteinian Framework}},\ }\href
  {https://doi.org/10.1103/PhysRevD.80.122003} {\bibfield  {journal} {\bibinfo
  {journal} {Phys. Rev. D}\ }\textbf {\bibinfo {volume} {80}},\ \bibinfo
  {pages} {122003} (\bibinfo {year} {2009})},\ \Eprint
  {https://arxiv.org/abs/0909.3328} {arXiv:0909.3328 [gr-qc]} \BibitemShut
  {NoStop}%
\bibitem [{\citenamefont {Gursel}\ and\ \citenamefont
  {Tinto}(1989)}]{Guersel:1989th}%
  \BibitemOpen
  \bibfield  {author} {\bibinfo {author} {\bibfnamefont {Y.}~\bibnamefont
  {Gursel}}\ and\ \bibinfo {author} {\bibfnamefont {M.}~\bibnamefont {Tinto}},\
  }\bibfield  {title} {\bibinfo {title} {{Near optimal solution to the inverse
  problem for gravitational wave bursts}},\ }\href
  {https://doi.org/10.1103/PhysRevD.40.3884} {\bibfield  {journal} {\bibinfo
  {journal} {Phys. Rev. D}\ }\textbf {\bibinfo {volume} {40}},\ \bibinfo
  {pages} {3884} (\bibinfo {year} {1989})}\BibitemShut {NoStop}%
\bibitem [{\citenamefont {Chatterji}\ \emph {et~al.}(2006)\citenamefont
  {Chatterji}, \citenamefont {Lazzarini}, \citenamefont {Stein}, \citenamefont
  {Sutton}, \citenamefont {Searle},\ and\ \citenamefont
  {Tinto}}]{Chatterji:2006nh}%
  \BibitemOpen
  \bibfield  {author} {\bibinfo {author} {\bibfnamefont {S.}~\bibnamefont
  {Chatterji}}, \bibinfo {author} {\bibfnamefont {A.}~\bibnamefont
  {Lazzarini}}, \bibinfo {author} {\bibfnamefont {L.}~\bibnamefont {Stein}},
  \bibinfo {author} {\bibfnamefont {P.~J.}\ \bibnamefont {Sutton}}, \bibinfo
  {author} {\bibfnamefont {A.}~\bibnamefont {Searle}},\ and\ \bibinfo {author}
  {\bibfnamefont {M.}~\bibnamefont {Tinto}},\ }\bibfield  {title} {\bibinfo
  {title} {{Coherent network analysis technique for discriminating
  gravitational-wave bursts from instrumental noise}},\ }\href
  {https://doi.org/10.1103/PhysRevD.74.082005} {\bibfield  {journal} {\bibinfo
  {journal} {Phys. Rev. D}\ }\textbf {\bibinfo {volume} {74}},\ \bibinfo
  {pages} {082005} (\bibinfo {year} {2006})},\ \Eprint
  {https://arxiv.org/abs/gr-qc/0605002} {arXiv:gr-qc/0605002} \BibitemShut
  {NoStop}%
\bibitem [{\citenamefont {Yunes}\ and\ \citenamefont
  {Siemens}(2013)}]{Yunes:2013dva}%
  \BibitemOpen
  \bibfield  {author} {\bibinfo {author} {\bibfnamefont {N.}~\bibnamefont
  {Yunes}}\ and\ \bibinfo {author} {\bibfnamefont {X.}~\bibnamefont
  {Siemens}},\ }\bibfield  {title} {\bibinfo {title} {{Gravitational-Wave Tests
  of General Relativity with Ground-Based Detectors and Pulsar
  Timing-Arrays}},\ }\href {https://doi.org/10.12942/lrr-2013-9} {\bibfield
  {journal} {\bibinfo  {journal} {Living Rev. Rel.}\ }\textbf {\bibinfo
  {volume} {16}},\ \bibinfo {pages} {9} (\bibinfo {year} {2013})},\ \Eprint
  {https://arxiv.org/abs/1304.3473} {arXiv:1304.3473 [gr-qc]} \BibitemShut
  {NoStop}%
\bibitem [{\citenamefont {Isi}\ and\ \citenamefont
  {Weinstein}(2017)}]{Isi:2017fbj}%
  \BibitemOpen
  \bibfield  {author} {\bibinfo {author} {\bibfnamefont {M.}~\bibnamefont
  {Isi}}\ and\ \bibinfo {author} {\bibfnamefont {A.~J.}\ \bibnamefont
  {Weinstein}},\ }\bibfield  {title} {\bibinfo {title} {{Probing gravitational
  wave polarizations with signals from compact binary coalescences}},\
  }\href@noop {} {\  (\bibinfo {year} {2017})},\ \Eprint
  {https://arxiv.org/abs/1710.03794} {arXiv:1710.03794 [gr-qc]} \BibitemShut
  {NoStop}%
\bibitem [{\citenamefont {Chatziioannou}\ \emph {et~al.}(2021)\citenamefont
  {Chatziioannou}, \citenamefont {Isi}, \citenamefont {Haster},\ and\
  \citenamefont {Littenberg}}]{Chatziioannou:2021mij}%
  \BibitemOpen
  \bibfield  {author} {\bibinfo {author} {\bibfnamefont {K.}~\bibnamefont
  {Chatziioannou}}, \bibinfo {author} {\bibfnamefont {M.}~\bibnamefont {Isi}},
  \bibinfo {author} {\bibfnamefont {C.-J.}\ \bibnamefont {Haster}},\ and\
  \bibinfo {author} {\bibfnamefont {T.~B.}\ \bibnamefont {Littenberg}},\
  }\bibfield  {title} {\bibinfo {title} {{Morphology-independent test of the
  mixed polarization content of transient gravitational wave signals}},\ }\href
  {https://doi.org/10.1103/PhysRevD.104.044005} {\bibfield  {journal} {\bibinfo
   {journal} {Phys. Rev. D}\ }\textbf {\bibinfo {volume} {104}},\ \bibinfo
  {pages} {044005} (\bibinfo {year} {2021})},\ \Eprint
  {https://arxiv.org/abs/2105.01521} {arXiv:2105.01521 [gr-qc]} \BibitemShut
  {NoStop}%
\bibitem [{\citenamefont {Hansen}\ \emph {et~al.}(2015)\citenamefont {Hansen},
  \citenamefont {Yunes},\ and\ \citenamefont {Yagi}}]{Hansen:2014ewa}%
  \BibitemOpen
  \bibfield  {author} {\bibinfo {author} {\bibfnamefont {D.}~\bibnamefont
  {Hansen}}, \bibinfo {author} {\bibfnamefont {N.}~\bibnamefont {Yunes}},\ and\
  \bibinfo {author} {\bibfnamefont {K.}~\bibnamefont {Yagi}},\ }\bibfield
  {title} {\bibinfo {title} {{Projected Constraints on Lorentz-Violating
  Gravity with Gravitational Waves}},\ }\href
  {https://doi.org/10.1103/PhysRevD.91.082003} {\bibfield  {journal} {\bibinfo
  {journal} {Phys. Rev. D}\ }\textbf {\bibinfo {volume} {91}},\ \bibinfo
  {pages} {082003} (\bibinfo {year} {2015})},\ \Eprint
  {https://arxiv.org/abs/1412.4132} {arXiv:1412.4132 [gr-qc]} \BibitemShut
  {NoStop}%
\bibitem [{\citenamefont {Lin}\ \emph {et~al.}(2019)\citenamefont {Lin},
  \citenamefont {Zhao}, \citenamefont {Zhang}, \citenamefont {Liu},
  \citenamefont {Wang}, \citenamefont {Zhang}, \citenamefont {Zhang},
  \citenamefont {Zhao}, \citenamefont {Zhu},\ and\ \citenamefont
  {Wang}}]{Lin:2018ken}%
  \BibitemOpen
  \bibfield  {author} {\bibinfo {author} {\bibfnamefont {K.}~\bibnamefont
  {Lin}}, \bibinfo {author} {\bibfnamefont {X.}~\bibnamefont {Zhao}}, \bibinfo
  {author} {\bibfnamefont {C.}~\bibnamefont {Zhang}}, \bibinfo {author}
  {\bibfnamefont {T.}~\bibnamefont {Liu}}, \bibinfo {author} {\bibfnamefont
  {B.}~\bibnamefont {Wang}}, \bibinfo {author} {\bibfnamefont {S.}~\bibnamefont
  {Zhang}}, \bibinfo {author} {\bibfnamefont {X.}~\bibnamefont {Zhang}},
  \bibinfo {author} {\bibfnamefont {W.}~\bibnamefont {Zhao}}, \bibinfo {author}
  {\bibfnamefont {T.}~\bibnamefont {Zhu}},\ and\ \bibinfo {author}
  {\bibfnamefont {A.}~\bibnamefont {Wang}},\ }\bibfield  {title} {\bibinfo
  {title} {{Gravitational waveforms, polarizations, response functions, and
  energy losses of triple systems in Einstein-aether theory}},\ }\href
  {https://doi.org/10.1103/PhysRevD.99.023010} {\bibfield  {journal} {\bibinfo
  {journal} {Phys. Rev. D}\ }\textbf {\bibinfo {volume} {99}},\ \bibinfo
  {pages} {023010} (\bibinfo {year} {2019})},\ \Eprint
  {https://arxiv.org/abs/1810.07707} {arXiv:1810.07707 [astro-ph.GA]}
  \BibitemShut {NoStop}%
\bibitem [{\citenamefont {Foster}(2007)}]{Foster:2007gr}%
  \BibitemOpen
  \bibfield  {author} {\bibinfo {author} {\bibfnamefont {B.~Z.}\ \bibnamefont
  {Foster}},\ }\bibfield  {title} {\bibinfo {title} {{Strong field effects on
  binary systems in Einstein-aether theory}},\ }\href
  {https://doi.org/10.1103/PhysRevD.76.084033} {\bibfield  {journal} {\bibinfo
  {journal} {Phys. Rev. D}\ }\textbf {\bibinfo {volume} {76}},\ \bibinfo
  {pages} {084033} (\bibinfo {year} {2007})},\ \Eprint
  {https://arxiv.org/abs/0706.0704} {arXiv:0706.0704 [gr-qc]} \BibitemShut
  {NoStop}%
\bibitem [{\citenamefont {Jacobson}(2014)}]{Jacobson:2013xta}%
  \BibitemOpen
  \bibfield  {author} {\bibinfo {author} {\bibfnamefont {T.}~\bibnamefont
  {Jacobson}},\ }\bibfield  {title} {\bibinfo {title} {{Undoing the twist: The
  Ho\v{r}ava limit of Einstein-aether theory}},\ }\href
  {https://doi.org/10.1103/PhysRevD.89.081501} {\bibfield  {journal} {\bibinfo
  {journal} {Phys. Rev. D}\ }\textbf {\bibinfo {volume} {89}},\ \bibinfo
  {pages} {081501(R)} (\bibinfo {year} {2014})},\ \Eprint
  {https://arxiv.org/abs/1310.5115} {arXiv:1310.5115 [gr-qc]} \BibitemShut
  {NoStop}%
\bibitem [{\citenamefont {Berti}\ \emph {et~al.}(2015)\citenamefont {Berti}
  \emph {et~al.}}]{Berti:2015itd}%
  \BibitemOpen
  \bibfield  {author} {\bibinfo {author} {\bibfnamefont {E.}~\bibnamefont
  {Berti}} \emph {et~al.},\ }\bibfield  {title} {\bibinfo {title} {{Testing
  General Relativity with Present and Future Astrophysical Observations}},\
  }\href {https://doi.org/10.1088/0264-9381/32/24/243001} {\bibfield  {journal}
  {\bibinfo  {journal} {Class. Quant. Grav.}\ }\textbf {\bibinfo {volume}
  {32}},\ \bibinfo {pages} {243001} (\bibinfo {year} {2015})},\ \Eprint
  {https://arxiv.org/abs/1501.07274} {arXiv:1501.07274 [gr-qc]} \BibitemShut
  {NoStop}%
\bibitem [{\citenamefont {Arun}\ \emph {et~al.}(2022)\citenamefont {Arun} \emph
  {et~al.}}]{LISA:2022kgy}%
  \BibitemOpen
  \bibfield  {author} {\bibinfo {author} {\bibfnamefont {K.~G.}\ \bibnamefont
  {Arun}} \emph {et~al.} (\bibinfo {collaboration} {LISA}),\ }\bibfield
  {title} {\bibinfo {title} {{New horizons for fundamental physics with
  LISA}},\ }\href {https://doi.org/10.1007/s41114-022-00036-9} {\bibfield
  {journal} {\bibinfo  {journal} {Living Rev. Rel.}\ }\textbf {\bibinfo
  {volume} {25}},\ \bibinfo {pages} {4} (\bibinfo {year} {2022})},\ \Eprint
  {https://arxiv.org/abs/2205.01597} {arXiv:2205.01597 [gr-qc]} \BibitemShut
  {NoStop}%
\bibitem [{\citenamefont {Buonanno}\ \emph {et~al.}(2009)\citenamefont
  {Buonanno}, \citenamefont {Iyer}, \citenamefont {Ochsner}, \citenamefont
  {Pan},\ and\ \citenamefont {Sathyaprakash}}]{Buonanno:2009zt}%
  \BibitemOpen
  \bibfield  {author} {\bibinfo {author} {\bibfnamefont {A.}~\bibnamefont
  {Buonanno}}, \bibinfo {author} {\bibfnamefont {B.~R.}\ \bibnamefont {Iyer}},
  \bibinfo {author} {\bibfnamefont {E.}~\bibnamefont {Ochsner}}, \bibinfo
  {author} {\bibfnamefont {Y.}~\bibnamefont {Pan}},\ and\ \bibinfo {author}
  {\bibfnamefont {B.~S.}\ \bibnamefont {Sathyaprakash}},\ }\bibfield  {title}
  {\bibinfo {title} {{Comparison of post-Newtonian templates for compact binary
  inspiral signals in gravitational-wave detectors}},\ }\href
  {https://doi.org/10.1103/PhysRevD.80.084043} {\bibfield  {journal} {\bibinfo
  {journal} {Phys. Rev. D}\ }\textbf {\bibinfo {volume} {80}},\ \bibinfo
  {pages} {084043} (\bibinfo {year} {2009})},\ \Eprint
  {https://arxiv.org/abs/0907.0700} {arXiv:0907.0700 [gr-qc]} \BibitemShut
  {NoStop}%
\bibitem [{\citenamefont {Blanchet}\ \emph {et~al.}(2023)\citenamefont
  {Blanchet}, \citenamefont {Faye}, \citenamefont {Henry}, \citenamefont
  {Larrouturou},\ and\ \citenamefont {Trestini}}]{Blanchet:2023bwj}%
  \BibitemOpen
  \bibfield  {author} {\bibinfo {author} {\bibfnamefont {L.}~\bibnamefont
  {Blanchet}}, \bibinfo {author} {\bibfnamefont {G.}~\bibnamefont {Faye}},
  \bibinfo {author} {\bibfnamefont {Q.}~\bibnamefont {Henry}}, \bibinfo
  {author} {\bibfnamefont {F.}~\bibnamefont {Larrouturou}},\ and\ \bibinfo
  {author} {\bibfnamefont {D.}~\bibnamefont {Trestini}},\ }\bibfield  {title}
  {\bibinfo {title} {{Gravitational-Wave Phasing of Quasi-Circular Compact
  Binary Systems to the Fourth-and-a-Half post-Newtonian Order}},\ }\href@noop
  {} {\  (\bibinfo {year} {2023})},\ \Eprint {https://arxiv.org/abs/2304.11185}
  {arXiv:2304.11185 [gr-qc]} \BibitemShut {NoStop}%
\bibitem [{\citenamefont {Moore}\ and\ \citenamefont
  {Nelson}(2001)}]{Moore:2001bv}%
  \BibitemOpen
  \bibfield  {author} {\bibinfo {author} {\bibfnamefont {G.~D.}\ \bibnamefont
  {Moore}}\ and\ \bibinfo {author} {\bibfnamefont {A.~E.}\ \bibnamefont
  {Nelson}},\ }\bibfield  {title} {\bibinfo {title} {{Lower bound on the
  propagation speed of gravity from gravitational Cherenkov radiation}},\
  }\href {https://doi.org/10.1088/1126-6708/2001/09/023} {\bibfield  {journal}
  {\bibinfo  {journal} {JHEP}\ }\textbf {\bibinfo {volume} {09}},\ \bibinfo
  {pages} {023}},\ \Eprint {https://arxiv.org/abs/hep-ph/0106220}
  {arXiv:hep-ph/0106220} \BibitemShut {NoStop}%
\bibitem [{\citenamefont {Kiyota}\ and\ \citenamefont
  {Yamamoto}(2015)}]{Kiyota:2015dla}%
  \BibitemOpen
  \bibfield  {author} {\bibinfo {author} {\bibfnamefont {S.}~\bibnamefont
  {Kiyota}}\ and\ \bibinfo {author} {\bibfnamefont {K.}~\bibnamefont
  {Yamamoto}},\ }\bibfield  {title} {\bibinfo {title} {{Constraint on modified
  dispersion relations for gravitational waves from gravitational Cherenkov
  radiation}},\ }\href {https://doi.org/10.1103/PhysRevD.92.104036} {\bibfield
  {journal} {\bibinfo  {journal} {Phys. Rev. D}\ }\textbf {\bibinfo {volume}
  {92}},\ \bibinfo {pages} {104036} (\bibinfo {year} {2015})},\ \Eprint
  {https://arxiv.org/abs/1509.00610} {arXiv:1509.00610 [gr-qc]} \BibitemShut
  {NoStop}%
\bibitem [{\citenamefont {Gong}\ and\ \citenamefont
  {Hou}(2018)}]{Gong:2017bru}%
  \BibitemOpen
  \bibfield  {author} {\bibinfo {author} {\bibfnamefont {Y.}~\bibnamefont
  {Gong}}\ and\ \bibinfo {author} {\bibfnamefont {S.}~\bibnamefont {Hou}},\
  }\bibfield  {title} {\bibinfo {title} {{Gravitational Wave Polarizations in
  $f(R)$ Gravity and Scalar-Tensor Theory}},\ }\href
  {https://doi.org/10.1051/epjconf/201816801003} {\bibfield  {journal}
  {\bibinfo  {journal} {EPJ Web Conf.}\ }\textbf {\bibinfo {volume} {168}},\
  \bibinfo {pages} {01003} (\bibinfo {year} {2018})},\ \Eprint
  {https://arxiv.org/abs/1709.03313} {arXiv:1709.03313 [gr-qc]} \BibitemShut
  {NoStop}%
\bibitem [{\citenamefont {Tachinami}\ \emph {et~al.}(2021)\citenamefont
  {Tachinami}, \citenamefont {Tonosaki},\ and\ \citenamefont
  {Sendouda}}]{Tachinami:2021jnf}%
  \BibitemOpen
  \bibfield  {author} {\bibinfo {author} {\bibfnamefont {T.}~\bibnamefont
  {Tachinami}}, \bibinfo {author} {\bibfnamefont {S.}~\bibnamefont
  {Tonosaki}},\ and\ \bibinfo {author} {\bibfnamefont {Y.}~\bibnamefont
  {Sendouda}},\ }\bibfield  {title} {\bibinfo {title} {{Gravitational-wave
  polarizations in generic linear massive gravity and generic higher-curvature
  gravity}},\ }\href {https://doi.org/10.1103/PhysRevD.103.104037} {\bibfield
  {journal} {\bibinfo  {journal} {Phys. Rev. D}\ }\textbf {\bibinfo {volume}
  {103}},\ \bibinfo {pages} {104037} (\bibinfo {year} {2021})},\ \Eprint
  {https://arxiv.org/abs/2102.05540} {arXiv:2102.05540 [gr-qc]} \BibitemShut
  {NoStop}%
\end{thebibliography}%

\end{document}